\documentclass[superscriptaddress, onecolumn, amsmath, amssymb, aps, prx]{revtex4-2}

\usepackage{graphicx}
\usepackage{dcolumn}
\usepackage{bm}
\usepackage{soul,xcolor}
\usepackage[normalem]{ulem}
\newcommand{\colorsout}[2]{%
  \bgroup
  \color{#1}
  \ULdepth=-.5ex 
  \sout{\textcolor{black}{#2}}
  \egroup
}

\begin{document}
\setstcolor{red}
\title{\textbf{Non-Hermitian Spectral Topology Shaped by Memory} 
}%

\author{Hao Pan}
\thanks{These authors contributed equally.}
\affiliation{Department of Physics, City University of Hong Kong, Kowloon, Hong Kong, China}

\author{Bo Xue}
\thanks{These authors contributed equally.}
\affiliation{Department of Electrical Engineering, City University of Hong Kong, Kowloon, Hong Kong, China}

\author{Wanyue Xiao}
\affiliation{Department of Physics, City University of Hong Kong, Kowloon, Hong Kong, China}

\author{Qiuyang Li}
\affiliation{Department of Electrical Engineering, City University of Hong Kong, Kowloon, Hong Kong, China}

\author{Yuqiong Cheng}
\affiliation{Department of Physics, City University of Hong Kong, Kowloon, Hong Kong, China}

\author{Yumeng Wang}
\affiliation{Department of Electrical Engineering, City University of Hong Kong, Kowloon, Hong Kong, China}

\author{Kun Ding}
\affiliation{Department of Physics, State Key Laboratory of Surface Physics, and Key Laboratory of Micro and Nano Photonic Structures (Ministry of Education), Fudan University, Shanghai, China}

\author{Alex M. H. Wong}
\email{alex.mh.wong@cityu.edu.hk}
\affiliation{Department of Electrical Engineering, City University of Hong Kong, Kowloon, Hong Kong, China}
\affiliation{State Key Laboratory of Terahertz and Millimeter Waves, City University of Hong Kong, Kowloon, Hong Kong, China}

\author{Shubo Wang}
\email{shubwang@cityu.edu.hk}
\affiliation{Department of Physics, City University of Hong Kong, Kowloon, Hong Kong, China}

\vspace{\baselineskip}
\renewcommand{\abstractname}{}
\begin{abstract}
Non-Hermitian spectral topology has transformed our understanding of bulk--boundary correspondence in open systems, yet its exploration has largely remained within the Markovian regime of instantaneous coupling. Here, we reveal how memory shapes non-Hermitian spectral topology by extending coupling asymmetry from space to time. Spatiotemporal nonreciprocity and memory-induced spectral multivaluedness give rise to nontrivial spectral topology on multi-sheeted Riemann surfaces, governed by both spatial connectivity and interaction history. Coherent interference among long-range memory channels can induce anomalous spectral winding and skin localization opposite to the dominant-coupling direction. Moreover, asymmetric retardation alone can generate non-Hermitian spectral topology and skin accumulation even in the absence of spatial coupling asymmetry. Channel dissipation further drives topological transitions that reverse the skin effect,  with a critical point at matched channel and lattice decay rates. We validate the anomalous skin effect and its dissipation-controlled reversal through full-wave simulations and microwave experiments. Our results establish time nonlocality as a fundamental degree of freedom in non-Hermitian topology, revealing phenomena inaccessible within the Markovian paradigm.
\end{abstract}

\maketitle

\section{Introduction}

Non-Hermitian Hamiltonians provide a fundamental framework for describing nonequilibrium dynamics in open systems. By extending band topology into the complex-energy plane, non-Hermiticity enables topological classifications based on point and line gaps \cite{bender2007making,ashida2020non,kawabata2019symmetry,Longhi2025Erratic}, giving rise to exotic phenomena without Hermitian counterparts \cite{bergholtz2021exceptional,ding2022non,hu2021knots,wang2021topological,patil2022measuring,sun2024photonic,lin2024observation,zhang2025algebraic}. A defining feature of this framework is the intimate connection between complex spectral topology and the boundary response. In particular, point-gap topology can produce nontrivial spectral winding and macroscopic accumulation of bulk eigenstates at system boundaries, leading to the non-Hermitian skin effect (NHSE) and a breakdown of conventional bulk-boundary correspondence \cite{yao2018edge,kunst2018biorthorgonal,yao2018non,okuma2018topological,zhang2020correspondence}. This boundary sensitivity is captured by non-Bloch band theory, which analytically continues the Bloch momentum into the complex plane and defines a generalized Brillouin zone (GBZ) that reconstructs the open-boundary spectrum \cite{yao2018edge,yokomizo2019non}. 

\textcolor{black}{Conventional studies of non-Hermitian topology typically employ effective Hamiltonians obtained by eliminating environmental degrees of freedom, leading to time-local Markovian dynamics in which environment-mediated couplings are regarded as instantaneous and memoryless \cite{weidemann2020topological,xiao2020non,ghatak2020observation,zhang2021acoustic,yu2020magnon,yu2022giant,yang2025non}. This approximation neglects the retardation and memory effects that naturally occur in realistic open systems}, including waveguide quantum electrodynamics, giant-atom architectures, and delay-coupled networks \cite{pichler2016photonic,ask2022non,andersson2019non,roccati2024controlling,grimsmo2015time,vodenkova2024continuous}, where interactions involve finite propagation times or structured environmental spectra. When environmental correlations persist over timescales comparable to the system dynamics, they cannot be adiabatically eliminated, giving rise to time-nonlocal dynamics characterized by a memory kernel and a frequency-dependent self-energy  \cite{breuer2002theory,chru2010non,rivas2014quantum,de2017dynamics,yu2024non,kuo2025non,Jana2026quantum}. However, how time-nonlocal memory fundamentally reshapes non-Hermitian topology and its associated NHSE remains largely unexplored.

Here, we reveal how memory reshapes non-Hermitian spectral topology beyond the Markovian regime. Retardation extends coupling asymmetry from space to time, giving rise to history-dependent spatiotemporal nonreciprocity, which, together with memory-induced spectral multivaluedness, generates nontrivial spectral topology on multi-sheeted Riemann surfaces. For long-range couplings with symmetric retardation, coherent superposition of multiple spatiotemporal pathways can induce anomalous topological winding and skin localization contrary to that expected from the dominant spatial coupling. In contrast, asymmetric retardation alone can generate non-Hermitian spectral topology and skin accumulation even in the absence of spatial coupling asymmetry. Channel dissipation provides an additional control parameter, driving non-Hermitian topological transitions accompanied by reversal of the skin effect. Experimentally validated in a coupled yttrium-iron-garnet (YIG)-array-waveguide microwave platform, these findings establish time nonlocality as a fundamental degree of freedom in non-Hermitian topology.

\section{Non-Markovian non-Hermitian spectral topology}
To illustrate this concept, we consider a one-dimensional (1D) time-translation-invariant lattice with time-retarded nearest-neighbor couplings [Fig. 1(a), upper panel], characterized by coupling amplitudes \(\Gamma_{L,R}\) and retardation times \(\tau_{L,R}\). The system can be viewed as an effective spatiotemporal lattice [Fig. 1(a), bottom panel], with couplings connecting sites at different spatial locations and times. This representation extends the conventional notion of nonreciprocity beyond purely spatial asymmetric hopping. Once retardation is introduced as an additional degree of freedom, reciprocity must be considered in the full spatiotemporal domain: it can be broken by either asymmetric coupling amplitudes (\(\Gamma_L \neq \Gamma_R\)) or asymmetric propagation delays (\(\tau_L \neq \tau_R\)). Specially, unequal delays map spatially opposite coupling onto inequivalent memory layers, corresponding to a shear deformation of the spatiotemporal lattice, thereby breaking the spatiotemporal inversion symmetry even when the corresponding spatial coupling amplitudes are identical. The resulting memory kernel is generally non-self-adjoint, providing an intrinsically non-Hermitian mechanism for nonreciprocal dynamics beyond the conventional paradigm of instantaneous asymmetric hopping.

\begin{figure}[ht]
    \centering
    \includegraphics[width=0.95\textwidth]{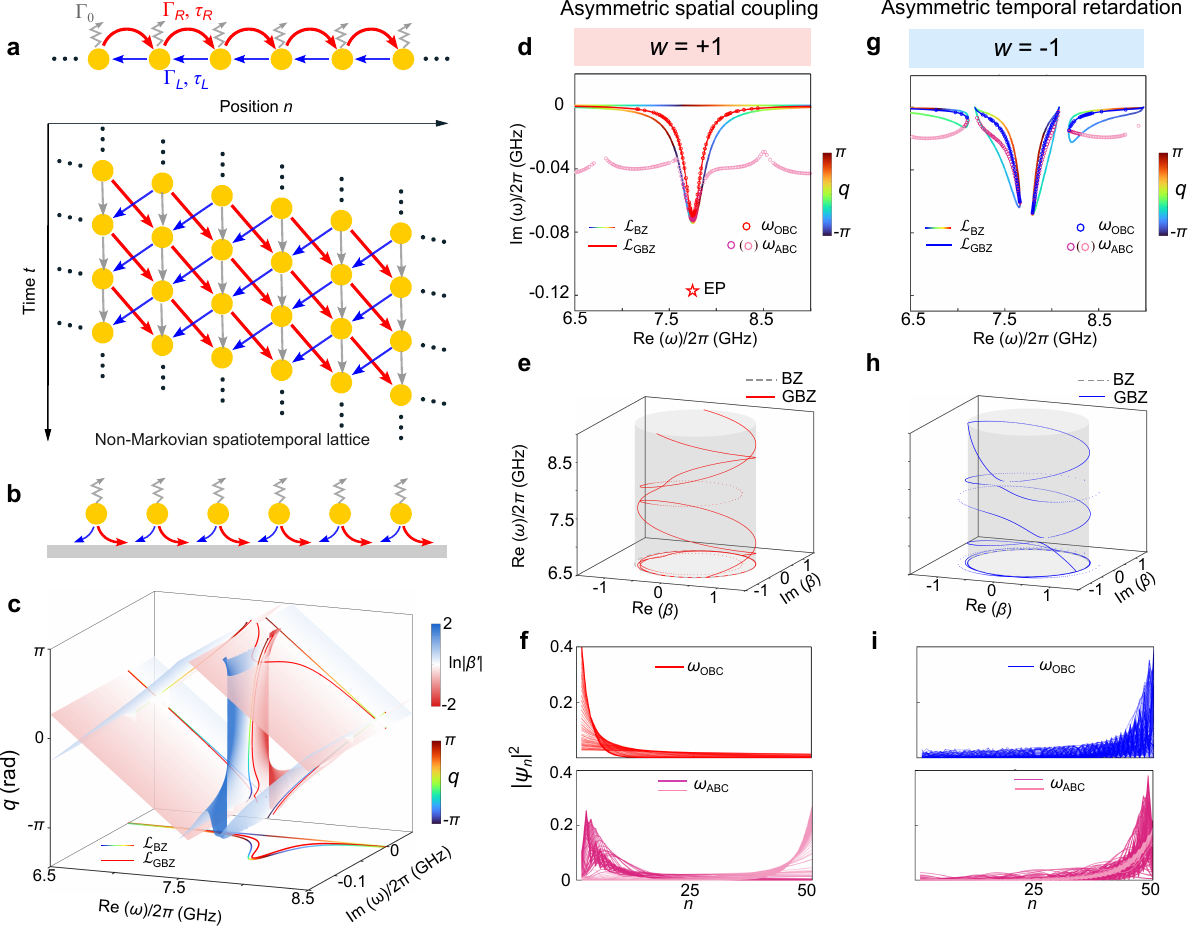}
    \caption{\textbf{Non-Markovian non-Hermitian 
    lattice model.} (a) Schematic of the non-Markovian spatiotemporal lattice with asymmetric spatial coupling and retardation time. (b) Schematic of a 1D lattice coupled through a continuum channel. (c) Complex spectra along BZ (\(\mathcal{L}_{\mathrm{BZ}}\)) and GBZ (\(\mathcal{L}_{\mathrm{GBZ}}\)), traversing the multi-sheeted Riemann surface due to analytical continuation to complex momentum \(q'\). The gradient color denotes \(\ln|\beta'|\) with \(\beta' = e^{iq'}\). Projected spectrum in the complex frequency plane for the lattices (d) with asymmetric spatial coupling and (g) with asymmetric temporal retardation. The red star denotes the EP for perfect unidirectional coupling under ABC. The GBZ contours for the lattices (e) with asymmetric spatial coupling and (h) with asymmetric temporal retardation. Eigenstate spatial distributions for a finite lattice \(N = 50\) (f) with asymmetric spatial coupling and (i) with asymmetric temporal retardation, under OBC (top) and ABC (bottom). For (c)-(f), we set \( \tilde{\omega}_0/2\pi = 7.75 - i0.08\) GHz, \(\Gamma_L/2\pi = 10^{-5}\) GHz, \(\Gamma_R/2\pi = 0.08\) GHz, and \(\tau_L = \tau_R = \tau = 0.59\) ns. For (g)-(i), we set \(\tilde{\omega}_0/2\pi = 7.75 - i0.08\) GHz, \(\Gamma_L/2\pi = \Gamma_R/2\pi = 0.08\) GHz, \(\tau_L = 0.16\) ns, and \(\tau_R = 0.96\) ns. The red and blue denote the left and right NHSE, respectively.}
\end{figure}

For a generic 1D spatiotemporal lattice with long-range couplings through a propagation continuum (e.g., waveguide) [Fig. 1(b)], the effective Hamiltonian is (see \textbf{Methods}),
\begin{equation}
H_{\mathrm{eff}}(\omega)=\sum_n\left(\omega_0-i \Gamma_t\right) a_n^{\dagger} a_n-i \Gamma \sum_n \sum_{m \neq n} e^{i \omega \tau|n-m|} a_n^{\dagger} a_m,
\end{equation}
where $a_n^\dagger (a_n)$ is the creation (annihilation) operator for the \(n\)th lattice site, and
$\Gamma_t=\Gamma_0+\left(\Gamma_L+\Gamma_R\right) / 2$ denotes the total damping including the intrinsic lattice damping $\Gamma_0$. The coupling strength $\Gamma$ and delay time between adjacent sites $\tau$ are direction dependent: \((\Gamma, \tau) = (\Gamma_L, \tau_L)\) for \(n < m\) and \((\Gamma, \tau) = (\Gamma_R, \tau_R)\) for \(n > m\), corresponding to left- and right-propagating channels, respectively. The directional delays are given by \(\tau_{L,R} = d/v_g^{L,R}\), where \(d\) is the lattice period and \(v_g^{L,R}\) are the group velocities of the left- and right-propagating channels. \textcolor{black}{Equation (1) shows that, at a single time slice, coupling retardation imparts a delay-dependent effective gain \(e^{-\omega_i\tau|n-m|}\) to decaying modes (\(\omega = \omega_r + i\omega_i\) with \(\omega_i<0\)), termed memory gain, stemming from the delayed arrival of signals from earlier, higher-amplitude states.} The corresponding time-domain equation of motion is
\begin{equation}
i \frac{d a_n(t)}{d t}=\left(\omega_0-i \Gamma_t\right) a_n(t)-i \Gamma \sum_{m \neq n} a_m(t-\tau|n-m|).
\end{equation}
The retardation term \(a_m(t-\tau|n-m|)\) explicitly accounts for the finite transit time of the coupling signal, thereby inducing non-Markovian dynamics where the time evolution of each site depends on the past states of its neighbors. Transforming Eq. (1) to the momentum space, we obtain the dispersion relation \textcolor{black}{ \(\omega = \omega_0-i\Gamma_t + f(\omega, \beta)\), where the coupling term \(f(\omega, \beta) = -i\Gamma_R e^{i\omega \tau_R}/(\beta - e^{i\omega \tau_R}) - i\Gamma_L e^{i\omega \tau_L}\beta/(1 - e^{i\omega \tau_L}\beta)\) with \(\beta=e^{iq}\) and \(q \in [-\pi, \pi)\).} This transcendental dispersion relation yields a multi-valued complex band structure that encodes the retardation-induced channel information. The corresponding GBZ is given by \(
\left| \beta_+ \right| = \left| \beta_- \right| = \left| e^{i\omega\Delta\tau}  \frac{\delta + i(\Gamma_0 - \Delta\Gamma/2)}{\delta + i(\Gamma_0 + \Delta\Gamma/2)} \right|^{1/2}
\) with \(\Delta\tau=\tau_R - \tau_L\), \( \Delta\Gamma = \Gamma_R - \Gamma_L
\), and \(
\delta = \omega - \omega_0\), which describes the bulk-eigenmode envelope of the spatiotemporal lattice under open boundary conditions (OBCs) in the thermodynamic limit \(N \rightarrow \infty\). Therefore, the spatial profile is determined by both the retardation asymmetry \(\Delta\tau\) and the coupling asymmetry \(\Delta\Gamma\), in stark contrast with conventional Markovian non-Hermitian systems.

We consider the nearly unidirectional right-coupling regime \(\Gamma_R/\Gamma_L \gg 1\) with symmetric retardation \(\tau_L = \tau_R = \tau\), resulting in symmetric memory gain. The multi-branched spectra \(\mathcal{L}_{\text{BZ}}\) under periodic boundary condition (PBC) resides on a multi-sheeted Riemann surface constructed via the analytic continuation of the momentum \(q' = q_r + iq_i\) [Fig. 1(c)], and forms a closed loop in the complex-frequency plane with a winding number \(w = +1\) [Fig. 1(d)]. According to the non-Markovian non-Bloch band theory (detailed in \textbf{SM}), the calculated GBZ also exhibits a multi-branched structure strictly confined within the unit circle \(|\beta| = 1\) [Fig. 1(e)], with the continuum spectrum \(\mathcal{L}_{\text{GBZ}}\) traversing the multi-sheeted Riemann surface and projected inside the PBC spectral loop \(\mathcal{L}_{\text{BZ}}\) [Figs. 1(c) and 1(d)]. Crucially, since the wavefunctions scale as \(\psi_n \propto \beta^n\), the condition \(|\beta| < 1\) dictates a macroscopic accumulation of bulk eigenstates at the left boundary, opposite to the direction of the dominant coupling. This anomalous skin localization still obeys the conventional non-Hermitian bulk-boundary correspondence, preserving the fundamental link between the topological winding number and the GBZ. Meanwhile, the emergent multi-branched spectral structure due to frequency-dependent self-energy fundamentally encapsulates the non-Markovian coupling history and environment-mediated propagation channel.

In the Markovian limit with retardation much shorter than the characteristic timescale (\(\tau \ll 1/\Gamma_t\)), the retardation effect reduces to a mere static phase factor, $a_m(t-\tau|n-m|)\rightarrow e^{i\omega_0\tau|n-m|}a_m(t)$, yielding a model governed by instantaneous, phase-modulated long-range couplings. The dispersion relation simplifies to $\omega = \tilde{\omega}_0  + \frac{\Gamma_R}{2}\cot((q_R - q)/2) + \frac{\Gamma_L}{2}\cot((q_L + q)/2)$, with \(\tilde{\omega}_0=\omega_0-i \Gamma_0\) and \(q_{L,R} = \omega_0 d/v_g^{L,R}\). Notably, its imaginary part becomes momentum-independent, \(\text{Im}(\omega(q)) = -\Gamma_0\), collapsing the complex spectrum onto an unbounded open line parallel to the real axis with ill-defined point-gap topology. Consequently, the instantaneous infinite-range coupling \textcolor{black}{without spatial scaling} constitutes a topological singularity that obliterates point-gap topology, irrespective of coupling asymmetry. Finite retardation lifts this singularity, thereby generating the nontrivial multi-branched spectral topology underlying the non-Markovian NHSE. 

To elucidate the origin of the anomalous non-Hermitian topological phase, we develop an effective Markovian picture by linearizing the frequency-dependent couplings of the non-Markovian model and mapping the retarded interactions onto renormalized instantaneous couplings. This effective picture correctly predicts the winding numbers for the cases in Fig. 1(d) (see \textbf{Methods}). For \(\Gamma_R/\Gamma_L \gg 1\) and \(\tau_L = \tau_R = \tau\), the linearized dispersion relation retaining the leading-order memory corrections reads \(\omega \approx \omega_0 - i\Gamma_t  + \frac{N(\beta_1)}{\beta_1 - \beta_2} \sum_{n=0}^{\infty} \beta_1^n \beta^{-(n+1)} + \frac{N(\beta_2)}{\beta_1 - \beta_2} \sum_{n=0}^{\infty} \beta_2^{-(n+1)} \beta^n\), where \(\beta_1\) and \(\beta_2\) \textcolor{black}{denote two poles arising from retardation and satisfying \(|\beta_1\beta_2|=1\) (\(|\beta_1|<1\) and \(|\beta_2|>1\)). The negative and positive powers of \(\beta\) correspond to the effective rightward and leftward coupling channels in the Markovian picture, respectively.} Interestingly, although the non-Markovian system is dominated by rightward coupling, the coefficients in the linearized dispersion reveal an effective larger leftward coupling than the rightward one in the projected Markovian space (see \textbf{Methods}). This counterintuitive reversal of the effective coupling asymmetry in the Markovian projection originates from memory-gain-enhanced long-range interactions, which regularize the singular infinite-range limit and drive the anomalous topological phase.

We consider the finite-size non-Markovian systems under two distinct boundary conditions: open boundary condition (OBC) and absorbing boundary condition (ABC). For OBCs, both the site array and the propagation channel are truncated and induce coherent boundary reflections. For ABCs, the site array is truncated, while the propagation channel (i.e., coupling channel) is completely absorbed at both boundaries without any reflection, effectively rendering the channel an infinite continuum. Figure 1(d) shows the discrete spectrum of a nearly unidirectional right-coupled system (\(N = 50\)) under ABCs (purple circles) and OBCs (red circles), with corresponding eigenstate profiles in Fig. 1(f). Under ABCs, the spectrum bifurcates: modes near the onsite resonance \(\widetilde{\omega}_0\) (dark purple) lie within \(\mathcal{L}_{\mathrm{BZ}}\) and exhibit topologically protected left-skin localization [Fig. 1(f), lower panel], whereas modes outside \(\mathcal{L}_{\mathrm{BZ}}\) (light purple) localize towards the right boundary. Unlike the conventional bipolar skin effect driven by twisted spectral winding \cite{zhang2021acoustic}, this bipolar skin localization emerges in the non-Markovian lattice despite a uniform coupling asymmetry. Distinct from Markovian systems, the discrete spectrum of the finite non-Markovian lattice gradually converges to \(\mathcal{L}_{\mathrm{GBZ}}\) as the system size increases (Fig. S4 in \textbf{SM}). Analytically, this convergence is dictated by the bulk root ratio, \(|\beta_+/\beta_-| = (C |\Gamma_R/\Gamma_L|)^{1/N}\), where $C\sim O(1)$. In the thermodynamic limit (\(N \to \infty\)), this relation naturally recovers the standard GBZ condition, \(|\beta_+| = |\beta_-|\). In the perfect unidirectional limit (\(\Gamma_L = 0\)), the Hamiltonian reduces to a right-cascaded Jordan chain \cite{wang2019arbitrary}. Its spectrum collapses to an \(N\)th-order exceptional point (EP) outside \(\mathcal{L}_{\mathrm{BZ}}\) at \(\omega_{\mathrm{EP}} = \omega_0 - i\Gamma_t\) [red star, Fig. 1(d)], with all fields strictly localized at the rightmost site. Remarkably, an arbitrarily weak leftward coupling lifts this high-order EP through coherent superposition of multiple feedback pathways within the spatiotemporal lattice. The resulting collective feedback can redirect retarded contributions towards the left side, where memory gain enhances their amplitude and produces a rightward-decaying spatial envelope across the lattice. Specifically, as \(\Gamma_L \to 0^+\), the spatial factor \(|\beta| \to e^{\omega_i \tau_L}\) (See \textbf{Methods}). Detuning from the lattice resonance frequency \(\widetilde{\omega}_0\) makes the right-cascaded dynamics dominant, restoring the rightward skin localization. 

Under OBC, the discrete spectrum closely follows \(\mathcal{L}_{\mathrm{GBZ}}\), and the corresponding eigenstates exhibit left-skin localization [Fig. 1(f), upper panel]. Here, the anomalous localization results from the combined effect of internal coherent retarded feedback and external boundary reflection.  Under perfect unidirectional right coupling, the internal feedback is absent; instead, boundary reflection returns the retarded field to the left end, effectively closing the propagation channel into a feedback loop. Consequently, the memory gain amplifies the returning field and imposes a rightward decay on the eigenmode spatial envelope, as captured by \(|\beta|^N|e^{i\omega N \tau_L}| = 1 \to |\beta| = e^{\omega_i \tau_L}\). Thus, the anomalous skin localization arises from the coherent amplification of retarded feedback, whether generated internally by site coupling or externally by boundary reflection, which reshapes the spatial envelope of the finite lattice.

Asymmetric retardation alone is sufficient to break spatiotemporal reciprocity and induce NHSE with no counterpart in conventional Markovian systems. To illustrate this, we consider the case with \(\tau_R > \tau_L\) and \(\Gamma_L = \Gamma_R\). Despite the spatial symmetry, the larger right retardation time induces a dominant effective right coupling, which drives the multi-branched PBC spectrum form multiple closed loops in the complex frequency plane, characterized by a winding number \(w = -1\) [Fig. 1(g)]. Consequently, the GBZ expands beyond the unit circle of \(|\beta| = 1\) [Fig. 1(h)], indicating a right-localization of the bulk eigenstates. For the finite system with \(N = 50\), the OBC eigenfrequencies \(\omega_{\text{OBC}}\) (blue circles) still follow the continuum spectrum \(\mathcal{L}_{\text{GBZ}}\) (blue line) defined by GBZ, and their corresponding states exhibit a right-localization profile [Fig. 1(i), upper panel]. Under ABC, the eigenfrequencies away from \(\tilde{\omega}_0\) still locate outside the \(\mathcal{L}_{\text{BZ}}\) loop, but all the eigenstates exhibit a right-skin localization [bottom panel of Fig. 1(i)],  in contrast to the case with asymmetric spatial couplings [Fig. 1(f)]. This contrast arises because symmetric spatial coupling preserves a spatially-balanced retarded feedback, which remains symmetric when external boundary reflections are removed under ABCs. In contrast, asymmetric retardation introduces an imbalance in memory gain: the longer rightward delay provides greater amplification for the field returning to the right boundary, thereby reshaping the macroscopic spatial envelope into a right-skin profile. Time nonlocality therefore constitutes an independent degree of freedom. More generally, it is the spatiotemporal nonreciprocity—rather than spatial coupling asymmetry alone—that fundamentally governs the non-Hermitian spectral topology of non-Markovian systems.

\section{Dissipation-driven topology and skin reversal}
Channel dissipation can alter the non-Hermitian spectral topology of the non-Markovian system. To elucidate, we incorporate a loss rate \(\Gamma_w\) into the propagation channel via the modified dispersion \(\omega_k = v_g |k| - i\Gamma_w\). Consequently, the retarded factor \(e^{i\omega\tau}\) in Eq. (1) can be modified as \(e^{(i\omega-\Gamma_w)\tau}\).
In the regime of \(\Gamma_R/\Gamma_L \gg 1\) with \(\tau_R = \tau_L\), the increasing \(\Gamma_w\) drives a topological phase transition, characterized by a reversal of the spectral winding of \(\mathcal{L}_{\mathrm{BZ}}\) from +1 to $-1$ [Fig. 2(a)]. Concurrently, the GBZ expands outward across the unit circle [Fig. 2(b)], dictating a reversal of the skin localization direction from left to right. On the other hand, for the cases with \(\tau_R > \tau_L\) and \(\Gamma_L = \Gamma_R\), the increasing \(\Gamma_w\) flips the winding number from $-1$ to +1 [Fig. 2(c)] and forces the GBZ to shrink inward across the BZ boundary [Fig. 2(d)], switching the localization from right to left. A critical phase transition point emerges exactly at \(\Gamma_w = \Gamma_0\), where \(\mathcal{L}_{\mathrm{BZ}}\) collapses onto an unbounded straight line at \(\mathrm{Im}(\omega) = -\Gamma_0\), rendering the point-gap topology ill-defined (see \textbf{Methods}).

\begin{figure}[t]
    \centering
    \includegraphics[width=\textwidth]{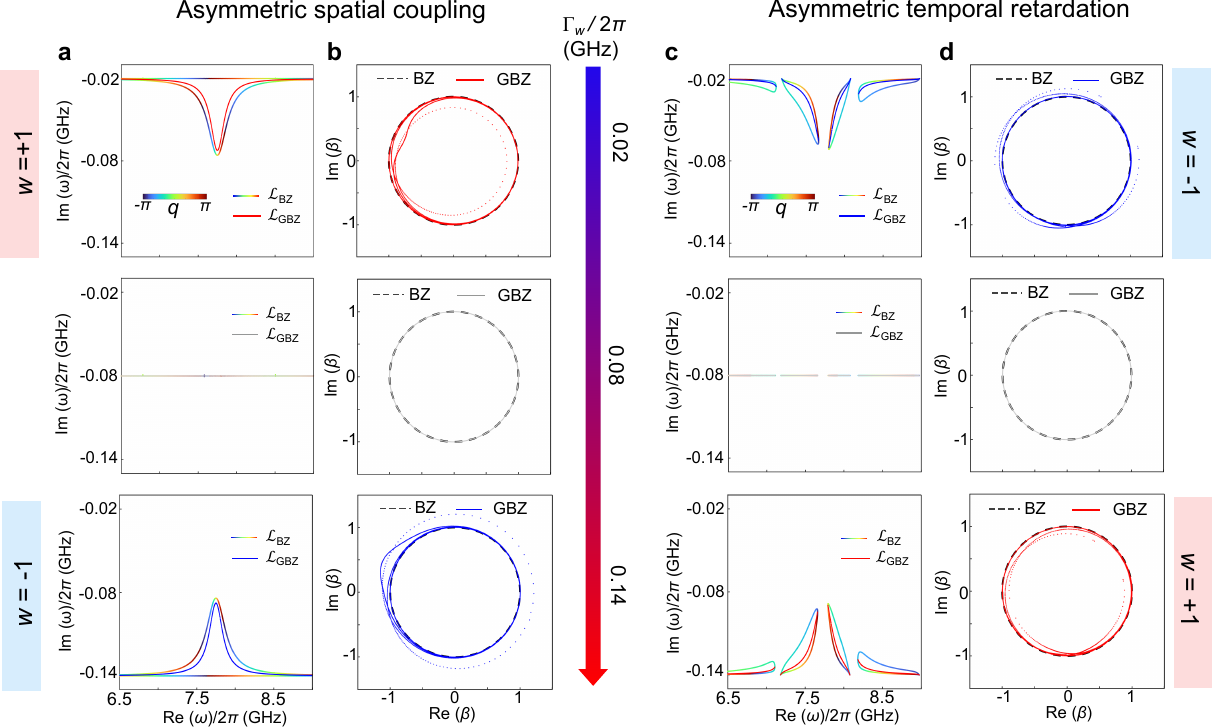}
    \caption{\textbf{Channel-dissipation-induced non-Hermitian topological transitions.} (a, b) Asymmetric spatial coupling regime (\(\Gamma_L/2\pi = 10^{-5}\) GHz, \(\Gamma_R/2\pi = 0.08\) GHz, \(\tau_L = \tau_R = 0.59\) ns, \(\tilde{\omega}_0/2\pi = 7.75 - i0.08\) GHz). (a) BZ spectrum \(\mathcal{L}_{\mathrm{BZ}}\) and GBZ continuous spectra \(\mathcal{L}_{\mathrm{GBZ}}\) for different channel loss \(\Gamma_w\) and (b) the corresponding GBZ contours in the complex momentum plane. Increasing \(\Gamma_w\) drives a topological transition of the winding number \(w\) from +1 to -1, while the GBZ expands outward across the standard BZ unit circle, marking a macroscopic crossover from the left to right NHSE. (c, d) Asymmetric temporal retardation regime (\(\tau_L = 0.16\) ns, \(\tau_R = 0.96\) ns, \(\Gamma_L/2\pi = \Gamma_R/2\pi = 0.08\) GHz, \(\tilde{\omega}_0/2\pi = 7.75 - i0.08\) GHz). The spectra evolution (c) and GBZ contours (d) exhibit opposite behaviors compared to spatially asymmetric case: increasing \(\Gamma_w\) flips \(w\) from -1 to +1 and forces the GBZ to shrink inward across the BZ boundary, indicating a right-to-left NHSE crossover. Solid red (blue) curves denote the GBZ and its associated spectrum located inside (outside) the standard BZ.}
\end{figure}

The phase transition originates from the memory kernel \(e^{(i\omega-\Gamma_w)\tau|n-m|}\). Through this kernel, the effective long-range coupling is determined by the imbalance between propagation attenuation \(\Gamma_w\) and collective-mode damping \(\omega_i\). Fundamentally, the lattice-channel hybridization bounds the system's effective decay rate between \(\Gamma_0\) and \(\Gamma_w\). At \(\Gamma_w = \Gamma_0\) where \(\omega_i = -\Gamma_w\), the memory-kernel amplitude satisfies \(e^{-(\Gamma_w+\omega_i)\tau|n-m|}=1\). The decay of the collective mode is then exactly balanced by that of the propagation channel, eliminating the spatial dependence of \textcolor{black}{memory-kernel amplitude} and restoring the singular infinite-range-coupling limit. Crossing this balance changes the effective weighting of the memory-mediated interactions, thereby driving the topological transition. The condition \(\Gamma_w = \Gamma_0\) thus marks a synchronization of the decay timescales of the lattice modes and the propagation channel. Although retardation persists at this point, the matched decay rates neutralize its distance-dependent amplitude renormalization, recovering the singular infinite-range limit with a closed point gap.

The accompanying reversal of the skin localization admits a direct physical interpretation in terms of the same gain–loss competition. \textcolor{black}{For asymmetric spatial coupling, when memory gain dominates propagation loss (\(\Gamma_0 > \Gamma_w\)), the net gain amplifies the retarded feedback field, driving the spatial envelope to decay along the dominant-coupling direction. Conversely, when channel dissipation dominates memory gain (\(\Gamma_0 < \Gamma_w\)), the retarded feedback field is instead heavily damped, restoring localization along the dominant coupling direction.} The same mechanism governs the effect of asymmetric retardation: in the memory-gain-dominated regime, a longer delay enhances the retarded field and drives localization towards the longer-delay direction, whereas in the channel-loss-dominated regime it produces stronger attenuation and drives localization toward the shorter-delay direction.

\section{Microwave realization}
We employ a microwave platform to demonstrate the non-Markovian NHSE under asymmetric spatial coupling and symmetric retardation [Fig. 3(a)]. The system consists of a 1D array of YIG cylinders coupled through a dielectric waveguide with relative permittivity \(\varepsilon_r = 12\). The YIG cylinders are biased by a static magnetic field \(B_0 = 0.1\) T along the -\(y\) direction. Under \(y\)-polarized plane-wave excitation, each cylinder directionally couples to the TE guided mode. This directionality can be characterized by the \(S\)-parameter contrast \(D=(|S_{21}|^2 - |S_{31}|^2)/(|S_{21}|^2 + |S_{31}|^2)\), with $D=+1 (D=-1)$ corresponding to perfect left (right) coupling. Full-wave simulation and experiment [Fig. 3(b)] consistently show that the directionality changes from $D=+1$ to $D=-1$ across the frequency range of interest, with the measured data (dashed line) agreeing well with full-wave simulations (solid line) [Fig. 3(c)]. This directionality reversal originates from opposite chiral magnetic quadrupole modes carrying opposite spin angular momentum (See \textbf{SM}), which selectively excite counterpropagating TE guided modes through spin-momentum locking \cite{bliokh2015quantum,cheng2023directional}. Consequently, the waveguide mediates directional, long-range interactions among the YIG cylinders, providing a physical realization of the nonreciprocal retarded couplings for the non-Markovian model.

\begin{figure}[t]
    \centering
    \includegraphics[width=\textwidth]{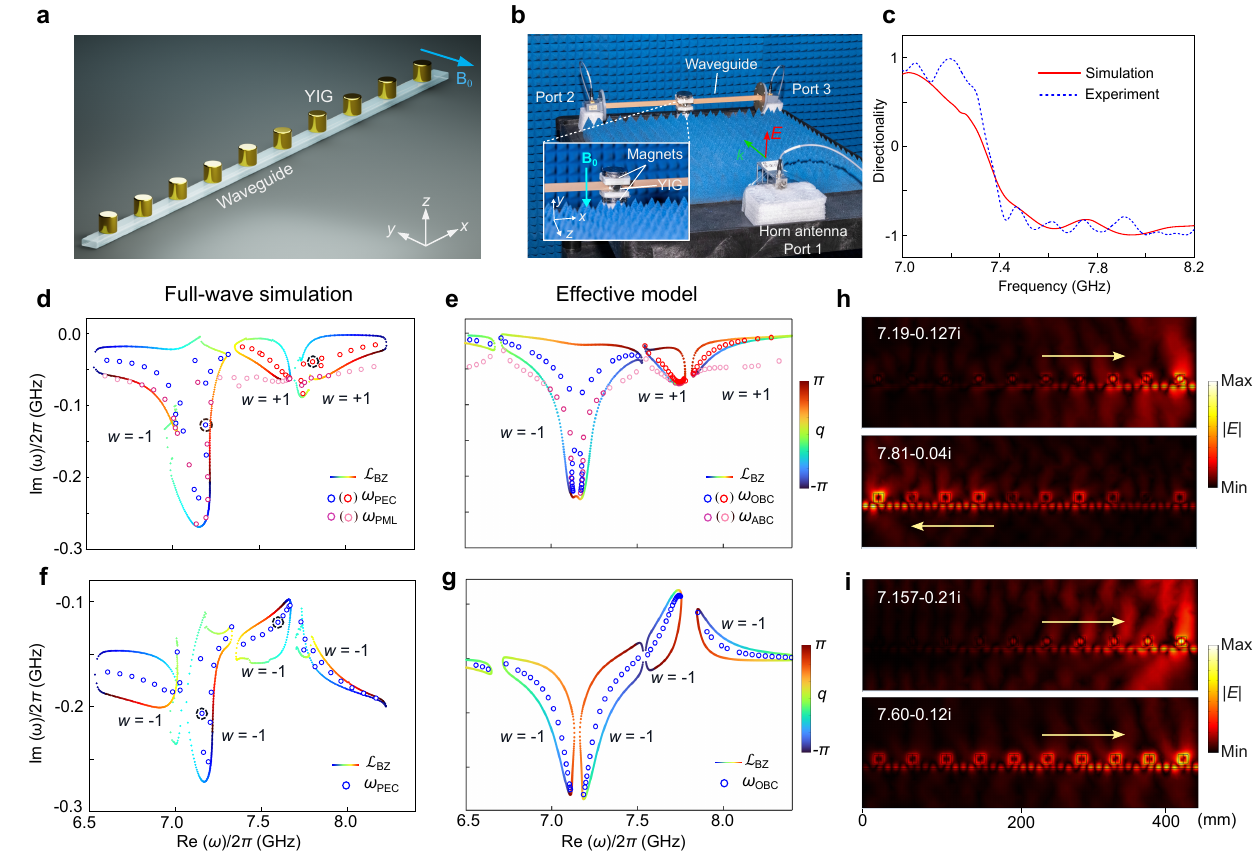}
    \caption{\textbf{Microwave realization of the non-Markovian NHSE.} (a) Schematic of 10-YIG-cylinder array (lattice constant \(d = 42.4\mathrm{mm}\)) side-coupled to a dielectric waveguide under an external magnetic bias \(B_0 = 0.1\) T. (b) Experimental setup for measuring the directional coupling of a YIG cylinder. (c) Directionality spectrum (\(D=(|S_{21}|^2 - |S_{31}|^2)/(|S_{21}|^2 + |S_{31}|^2)\)) for the single YIG cylinder-waveguide system under plane-wave excitation. (d, f) Simulated complex energy spectra without (\(\varepsilon_r = 12\)) and with (\(\varepsilon_r = 12 + 0.6i\)) intrinsic waveguide loss, respectively. (e, g) Analytical reproductions of (e) and (g). The model parameters \((\tilde{\omega}_0^{\pm}/2\pi, \Gamma_L^{\pm}/2\pi, \Gamma_R^{\pm}/2\pi)\) are \((7.15 - i0.26, 0.1, 0.01)\) and \((7.75 - i0.08, 0.005, 0.1)\) GHz for the lossless waveguide (\(\Gamma_w = 0\)) in (e), and \((7.15 - i0.3, 0.1, 0.021)\) and \((7.75 - i0.08, 0.023, 0.1)\) GHz for the lossy waveguide (\(\Gamma_w/2\pi = 0.15\) GHz) in (g). Both cases assume a symmetric retardation time \(\tau_L = \tau_R = \tau = 0.59\) ns. Solid lines delineate the BZ spectrum (\(\mathcal{L}_{\mathrm{BZ}}\)). Blue (red) hollow and purple circles denote the eigenvalues under PEC (OBC) and PML (ABC), respectively. For the PEC (OBC) case, red and blue circles denote the states with winding numbers of +1 and -1. For the PML (ABC) case, dark and light purple circles further distinguish states localized inside and outside the \(\mathcal{L}_{\mathrm{BZ}}\) loops. (h, i) Simulated electric-field distributions for representative eigenstates of \(\omega_{\mathrm{PEC}}\) for the cases without (h) and with (i) waveguide intrinsic loss highlighted by black dashed circles in (d) and (f), respectively.}
\end{figure}

The simulated PBC spectrum \(\mathcal{L}_{\mathrm{BZ}}\) forms three distinct loops with winding numbers -1, +1, and +1 from left to right, respectively [Fig. 3(d)]. For a finite array of \(N=10\) cylinders, we impose perfect-electric-conductor (PEC) and perfect-matched-layer (PML) boundaries to realize OBCs and ABCs, respectively. Under PEC boundaries, the eigenfrequencies \(\omega_{\mathrm{PEC}}\) (blue and red circles) lie within the corresponding \(\mathcal{L}_{\mathrm{BZ}}\) loops. Under PML boundaries, the eigenfrequencies \(\omega_{\mathrm{PML}}\) (dark purple circles) near the cylinder resonances \(\tilde{\omega}_0^+\) and \(\tilde{\omega}_0^-\) remain inside the \(\mathcal{L}_{\mathrm{BZ}}\) loops, whereas the remaining modes (light purple circles) extend outside them [Fig. 3(d)]. These numerical results are quantitatively reproduced by the non-Markovian long-range-coupling model [Fig. 3(e)]. Figure 3(h) shows representative eigenmode electric-field distributions under PEC boundaries for eigenfrequencies enclosed by the \(\mathcal{L}_{\mathrm{BZ}}\) loops with winding numbers -1 and +1, respectively [black dashed circles in Fig. 3(d)]. The corresponding field profiles exhibit skin localization at the right and left boundaries, respectively. Notably, both skin modes emerge in directions opposite to their respective dominant couplings depicted in Fig. 3(c), directly demonstrating the anomalous NHSE predicted by the non-Markovian theory.

To verify the dissipation-driven topological transition, we introduce loss into the waveguide by setting \(\varepsilon_r = 12 + 0.6i\). Figure 3(f) shows the simulated complex spectra under PBCs (solid lines), together with the eigenfrequencies of the finite \(N=10\) lattice under PEC boundaries (circles). Compared with the lossless case in Fig. 3(d), the low-frequency spectral loop with \(w = -1\) splits into two loops while retaining the same winding, whereas the two \(w = +1\) loops at higher frequencies reverse their winding to \(w = -1\) (Minor spectral kinks due to reciprocal air-channel coupling vanish at larger lattice spacings; see \textbf{SM}). This dissipation-induced topological transition is quantitatively captured by the non-Markovian model [Fig. 3(g)] and can be understood from the competition between the dissipation rates of the cylinder modes and guided wave. The two cylinder modes, \(\tilde{\omega}_0^+/2\pi = 7.15 - 0.28i\) GHz and \(\tilde{\omega}_0^-/2\pi = 7.75 - 0.08i\) GHz, exhibit predominant leftward and rightward couplings, respectively. The waveguide dissipation is \(\Gamma_w/2\pi = 0.15\) GHz, which lies between their intrinsic decay rates. Therefore, the condition for the dissipation-driven transition is satisfied only for the \(\tilde{\omega}_0^-\)-dominated bands, reversing their winding from +1 to -1, while the topology of the \(\tilde{\omega}_0^+\)-dominated low-frequency bands remains unchanged. Correspondingly, the representative PEC eigenmodes [black dashed circles in Fig. 3(f)] enclosed by the resulting \(w=-1\) spectral loops exhibit the right-skin localization [Fig. 3(i)].

\begin{figure}[tb]
    \centering
    \includegraphics[width=\textwidth]{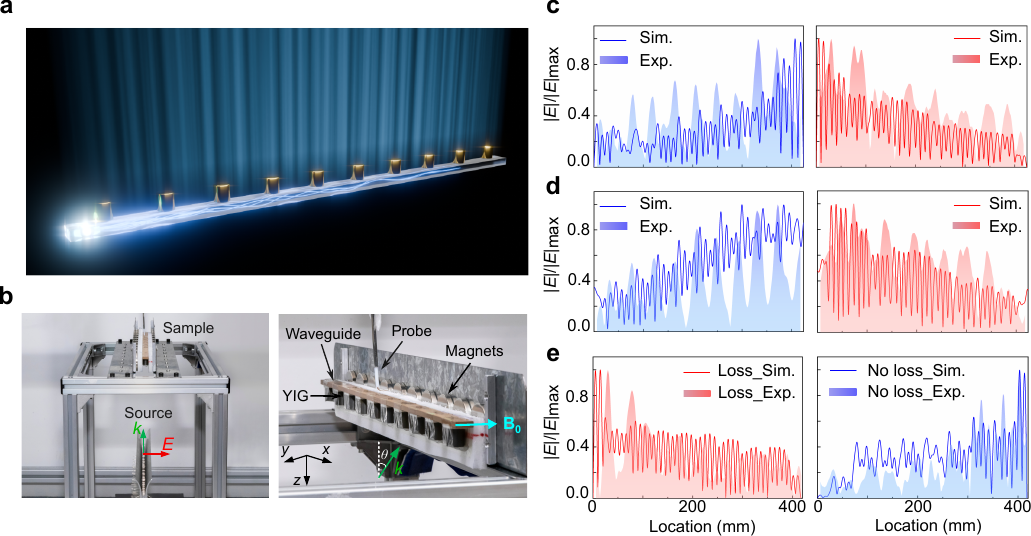}
    \caption{\textbf{Microwave realization of the non-Hermitian light funneling and dissipation-driven non-Hermitian topological transition.} (a) Schematic of light funneling function based on the NHSE. (b) A photo of the experimental setup. The system is excited by a plane wave via a meta-cavity antenna, with the electric field distribution mapped by a near-field electric probe. (c-e) Simulated (solid lines) and measured (shaded areas) normalized electric field distributions \(|E|/|E|_{\max}\) along the waveguide under a 0.1 T magnetic bias along the \(-y\)-direction. (c, d) Field spatial profiles for a lossless waveguide (\(\varepsilon_r = 12\)) coupled to a 10-YIG-cylinder array (\(d = 42.4\text{mm}\)) with (c) PEC and (d) PML boundaries. Left column displays right-localized skin states excited at 7.15 GHz (\(\theta = 0^\circ\) for PEC and \(-2^\circ\) for PML). Right column exhibits left-localized states at 7.81 GHz (\(\theta = -12^\circ\) for PEC and \(-8^\circ\) for PML). (e) Reversal of the excited skin effect at 7.76 GHz induced by waveguide loss in a 9-cylinder YIG array system with \(d = 46.4\text{ mm}\) for PEC boundary. The left and right columns correspond to the lossless (\(\varepsilon_r = 12\)) and lossy (\(\varepsilon_r = 12 + 0.4i\)) waveguide scenarios, with respective incidence angles of \(\theta = -20^\circ\) and \(\theta = 12^\circ\).}
\end{figure}

The microwave platform in Fig. 3(a) can enable selective free-space-to-waveguide conversion and directional energy routing, realizing a non-Hermitian light-funneling functionality that concentrates incident radiation toward one end of the waveguide [Fig. 4(a)]. To experimentally demonstrate the NHSE and the light-funneling functionality, we fabricate an array of ten YIG cylinders above a dielectric waveguide [Fig. 4(b)]. A static magnetic field of 0.1 T is applied along the \(-y\) direction using permanent magnets. A meta-cavity antenna is employed to generate a \(y\)-polarized plane wave incident in the \(z\)-\(x\) plane at an angle \(\theta\), \textcolor{black}{which matches the wave vector of the desired skin mode.} We use copper plates and microwave absorbers placed at both ends of waveguide to implement the PEC and PML boundaries, respectively. At 7.15 GHz, the measured field under PEC boundaries exhibits pronounced right-skin localization for normal incidence [left column of Fig. 4(c)]. Replacing the PEC boundaries with PMLs yields a similar right-localized field envelope [left column of Fig. 4(d)]. Conversely, tuning the frequency to 7.81 GHz abruptly shifts the accumulation, yielding a robust left-skin localization [right columns of Figs. 4(c) and 4(d)]. This excellent agreement with full-wave simulations unambiguously confirms the counterintuitive skin localization occurring opposite to the dominant-coupling direction, thereby validating the anomalous non-Markovian NHSE.

We further demonstrate the dissipation-driven non-Hermitian topological transition by introducing thin absorptive layers to the waveguide. For a finite array with PEC boundaries, this engineered dissipation induces a right-localized skin state at 7.76 GHz [right column of Fig. 4(e)]. Remarkably, removing these absorptive layers completely reverses the localization to the left at the exact same frequency [left column of Fig. 4(e)]. This stark contrast, perfectly captured by our simulations, provides definitive experimental proof that channel dissipation can drive the topological phase transition and reverse the skin-localization direction.

\section{Discussion}
In summary, we reveal how memory reshapes non-Hermitian spectral topology beyond the Markovian regime by introducing retardation as an intrinsic dynamical degree of freedom. Time-nonlocal interactions generate frequency-dependent couplings, rendering the eigenvalue problem nonlinear and the spectrum multivalued, with the hidden channel information unfolding onto multi-sheeted Riemann surfaces. The retardation extends coupling asymmetry from space to time, giving rise to spatiotemporal nonreciprocity. Together, spatiotemporal nonreciprocity and memory-induced spectral multivaluedness generate nontrivial spectral topology beyond the conventional Markovian description.

We show that memory fundamentally reshapes the skin effect and gives rise to counterintuitive phenomena. Under symmetric retardation, coherent retarded feedback can overcome the bare spatial coupling asymmetry, producing anomalous skin localization opposite to the dominant-coupling direction. Conversely, asymmetric retardation can independently generate the NHSE, even in the absence of spatial coupling asymmetry, through direction-dependent memory amplification. Channel dissipation provides an additional control parameter by suppressing memory gain and driving a topological transition when the decay rates of the propagation channel and lattice modes balance. At this critical point, the delay-dependent amplitude weighting of the memory channels vanishes, neutralizing the memory-induced topological renormalization and recovering the singular infinite-range limit with a closed point gap. We validate these predictions through full-wave simulations and microwave experiments, demonstrating the anomalous NHSE, dissipation-controlled localization reversal, and directional funneling of free-space radiation.

Our work establishes time nonlocality as a fundamental degree of freedom in non-Hermitian topology and extends nonreciprocity from a purely spatial concept to a spatiotemporal one. By revealing how memory, retardation, and dissipation collectively shape spectral topology and boundary localization, our results open a route towards non-Hermitian topological physics beyond the Markovian paradigm, with opportunities for controlling wave transport and dynamical responses across photonic, acoustic, and quantum platforms.

\nocite{*}

\bibliography{apssamp}

\clearpage 
\setcounter{equation}{0} 
\section*{Methods}
\renewcommand{\theequation}{M\arabic{equation}}
\subsection*{Non-Markovian Hamiltonian with long-range coupling}
We consider a 1D lattice of period \(d\) and different sites couple through a propagation continuum (e.g., waveguide), described by the Hamiltonian
\begin{equation}
H = \sum_n \tilde{\omega}_0 a_n^\dagger a_n + \frac{1}{2\pi} \int_{-\infty}^{+\infty} dk\, \omega_k c_k^\dagger c_k + \frac{1}{2\pi} \int_{-\infty}^{+\infty} dk \left( g_k e^{iknd} a_n^\dagger c_k + h.c. \right),
\end{equation}
where \(a_n^\dagger(a_n)\) and \(c_k^\dagger(c_k)\) are the creation (annihilation) operators for the \(n\)th site and the propagation wave with wavenumber \(k\), $\tilde{\omega}_0=\omega_0-i\Gamma_0$ is the onsite potential, and \(g_k\) characterizes the coupling between the sites and the propagation channel. We assume a linear dispersion of the continuum channel,
\begin{equation}
\omega_k = \begin{cases}
v_g^R k - i\Gamma_w, & k > 0 \\
-v_g^L k - i\Gamma_w, & k < 0
\end{cases},
\end{equation}
with right and left group velocity \(v_g^R\) and \(v_g^L\). By integrating out the continuum degrees of freedom of the propagation channel, we can map the full system onto an effective Hamiltonian for the lattice sites: \(H_{\text{eff}}(\omega) = H_0 + H_{\text{int}}(\omega)\). Here, \(H_0\) is the bare Hamiltonian, and \(H_{\text{int}}(\omega) = \sum_{n,m} J_{nm}(\omega) a_n^\dagger a_m\) is the non-Hermitian part that describes both coherent and dissipative site-site couplings mediated by the propagation channel, with the coupling strength \(J_{nm}(\omega) = \frac{1}{2\pi} \int_{-\infty}^{+\infty} dk \frac{|g_k|^2 e^{ikd(n-m)}}{\omega - \omega_k + i0^+}\). For simplification, we assume \(k\)-independent coupling constants, \(g_R\) and \(g_L\), for the right and left couplings, respectively. The frequency-domain effective Hamiltonian writes
\begin{equation}
H_{\mathrm{eff}}(\omega)=\sum_n\left(\omega_0-i \Gamma_t\right) a_n^{\dagger} a_n-i \Gamma \sum_n \sum_{m \neq n} e^{(i \omega-\Gamma_w) \tau|n-m|} a_n^{\dagger} a_m,
\end{equation}
where the coupling strength $\Gamma$ and unit retardation time $\tau$ are assigned by the propagation direction: \(\Gamma_R = |g_R|^2 / v_g^R\), \(\tau_R = d / v_g^R\) for right-propagating channel (\(m<n\)), and \(\Gamma_L = |g_L|^2 / v_g^L\), \(\tau_L = d / v_g^L\) for left-propagating channel (\(m>n\)), and the total dissipation \(\Gamma_t=\Gamma_0+\left(\Gamma_R+\Gamma_L\right) / 2\). The frequency-dependent phase factors in the effective Hamiltonian signify an inherently retarded channel-mediated interaction, a direct consequence of the finite propagation velocity of the channel mode. Inverse temporal Fourier transform of these factors to the time domain maps them onto delayed amplitudes, \(\frac{1}{2\pi} \int_{-\infty}^{+\infty} d\omega\, a_m(\omega) e^{i\omega \tau} e^{-i\omega t} = a_m(t - \tau)\), thus yielding
\begin{equation}
i \frac{d a_n(t)}{d t}=\left(\omega_0-i \Gamma_t\right) a_n(t)-i \Gamma \sum_{m \neq n} e^{-\Gamma_w\tau|n-m|} a_m(t-\tau|n-m|),
\end{equation}
 which explicitly captures the nonlocal-in-time nature of the coupling: the finite propagation delay dictates that the instantaneous evolution of any given site depends on the past states of the other sites, rendering the effective dynamics fundamentally non-Markovian.
 
 \textcolor{black}{By spatial Fourier transformation to momentum space, the Hamiltonian can be rewritten as \(H_{\text{eff}}(\omega) = \sum_q H(q, \omega) a_q^\dagger a_q\), with 
\begin{equation}
H(q, \omega) = \omega_0 -i\Gamma_t - i\Gamma_R \sum_{l=1}^{\infty} e^{(i\omega-\Gamma_w)l\tau_R} e^{-ilq} - i\Gamma_L \sum_{l=1}^{\infty} e^{(i\omega-\Gamma_w)l\tau_L} e^{ilq}.
\end{equation}
The self-consistent dispersion relation is (see \textbf{SM} for the detailed derivations)
\begin{equation}
\omega
=
\omega_0
-i\Gamma_t
-i\Gamma_R
\frac{e^{(i\omega-\Gamma_w)\tau_R}e^{-iq}}
{1-e^{(i\omega-\Gamma_w)\tau_R}e^{-iq}}
-i\Gamma_L
\frac{e^{(i\omega-\Gamma_w)\tau_L}e^{iq}}
{1-e^{(i\omega-\Gamma_w)\tau_L}e^{iq}} .
\end{equation}
}

\subsection*{Non-Markovian Hamiltonian of finite-size lattice}
We now consider a finite-size lattice with \(N\) sites. Under ABCs, the propagation continuum are completely dissipated at the boundaries without any reflection. The effective Hamiltonian can be expressed as
\begin{equation}
\begin{aligned}
H_{\text{eff}}^{\text{ABC}} &= \sum_{n=1}^N \left( \omega_0-i \Gamma_t \right) a_n^\dagger a_n - i\Gamma_R \sum_{1 \le m < n \le N} e^{(i\omega - \Gamma_w)(n-m)\tau_R} a_n^\dagger a_m \\
&\quad - i\Gamma_L \sum_{1 \le n < m \le N} e^{(i\omega - \Gamma_w)(m-n)\tau_L} a_n^\dagger a_m .
\end{aligned}
\end{equation}
In contrast,  OBCs cause complete wave reflections at the boundaries. Under this circumstance, the effective Hamiltonian is 
\begin{equation}
H_{\text{eff}}^{\text{OBC}}(\omega) = \sum_{n=1}^N \tilde{\omega}_0 a_n^\dagger a_n + \sum_{1 \le n,m \le N} J_{nm}^{\text{OBC}}(\omega) a_n^\dagger a_m,
\end{equation}
where the coupling \(J_{nm}^{\text{OBC}}(\omega)\), acting as the Green's function of the truncated continuum and capturing coherent feedback from multiple boundary reflections, is expressed as
\begin{equation}
\begin{aligned}
J_{nm}^{\text{OBC}}(\omega) &= -\frac{i}{1 - e^{(i\omega - \Gamma_w)(N+1)(\tau_R + \tau_L)}} \\
&\quad \times \begin{cases}
\begin{aligned}
&\Gamma_R e^{(i\omega - \Gamma_w)(n-m)\tau_R} + \Gamma_L e^{(i\omega - \Gamma_w)[m\tau_L + (N+1)\tau_R + (N+1-n)\tau_L]} \\
&- \Lambda e^{(i\omega - \Gamma_w)(m\tau_L + n\tau_R)} - \Lambda^* e^{(i\omega - \Gamma_w)[(N+1-m)\tau_R + (N+1-n)\tau_L]},
\end{aligned} & m < n \\
\begin{aligned}
&\frac{\Gamma_R + \Gamma_L}{2} \left( 1 + e^{(i\omega - \Gamma_w)(N+1)(\tau_R + \tau_L)} \right) \\
&- \Lambda e^{(i\omega - \Gamma_w)n(\tau_R + \tau_L)} - \Lambda^* e^{(i\omega - \Gamma_w)(N+1-n)(\tau_R + \tau_L)},
\end{aligned} & m = n \\
\begin{aligned}
&\Gamma_L e^{(i\omega - \Gamma_w)(m-n)\tau_L} + \Gamma_R e^{(i\omega - \Gamma_w)[(N+1-m+n)\tau_R + (N+1)\tau_L]} \\
&- \Lambda e^{(i\omega - \Gamma_w)(m\tau_L + n\tau_R)} - \Lambda^* e^{(i\omega - \Gamma_w)[(N+1-m)\tau_R + (N+1-n)\tau_L]},
\end{aligned} & m > n
\end{cases}
\end{aligned}
\end{equation}
with the cross-coupling amplitude \(\Lambda = g_L^* g_R / \sqrt{v_g^R v_g^L}\) and its complex conjugate \(\Lambda^*\). The detailed derivations can be found in \textbf{SM}.

\subsection*{Linear-approximation effective model}
We map the non-Markovian model to a Markovian picture, in which the temporally-retarded couplings are converted to renormalized instantaneous couplings. The non-Markovian nonlinear dispersion without channel dissipation (\(\Gamma_w = 0\)) can be simplified as \(\omega - \omega_0 = -i\Gamma_t + f(\omega, \beta)\), where the coupling term \(f(\omega, \beta) = -i\Gamma_R e^{i\omega \tau_R}/(\beta - e^{i\omega \tau_R}) - i\Gamma_L e^{i\omega \tau_L}\beta/(1 - e^{i\omega \tau_L}\beta)\) with \(\beta = e^{iq}\). Under the condition of \(|(\omega - \omega_0)\tau_{L,R}| < 1\), we apply Taylor expansion to \(f(\omega, \beta)\) at \(\omega = \omega_0\) and obtain the first-order linear-approximation dispersion relation
\begin{equation}
\omega - \omega_0 \approx \frac{ -i\Gamma_t - i\Gamma_Re^{i\omega_0 \tau_R}/(\beta - e^{i\omega_0 \tau_R}) - i\Gamma_Le^{i\omega_0 \tau_L}\beta/(1 - e^{i\omega_0 \tau_L}\beta)}{ 1 - \beta \left[\Gamma_R\tau_R e^{i\omega_0 \tau_R}/(\beta - e^{i\omega_0 \tau_R})^2 + \Gamma_L \tau_L e^{i\omega_0 \tau_L}/(1 - e^{i\omega_0 \tau_L}\beta)^2\right] }.
\end{equation}
When \(\tau_{L,R} \to 0\), the linearized dispersion reduces to the Markovian dispersion. Notably, this linear approximation is equivalent to the single Newton-iteration with the initial value \(\omega = \omega_0\), i.e., \(\omega_{\text{New}} = \omega_0 - g(\omega_0, \beta)/g'(\omega_0, \beta) = \omega_0 + (-i\Gamma_t + f(\omega_0, \beta))/(1 - f'(\omega_0, \beta))\) with \(g(\omega, \beta) = \omega_0 - i\Gamma_t + f(\omega, \beta) - \omega\).

For the nearly-unidirectional right-coupling case (\(\Gamma_R / \Gamma_L \gg 1\)) with symmetric retardation (\(\tau_L = \tau_R = \tau\)), Equation (M10) further simplifies to \(\omega(\beta) \approx \omega_0 + N(\beta)/D(\beta)\) with \(N(\beta) = -i\Gamma_t (\beta - e^{i\omega_0 \tau})^2 - i\Gamma_R e^{i\omega_0 \tau}(\beta - e^{i\omega_0 \tau})\) and \(D(\beta) = (\beta - e^{i\omega_0 \tau})^2 - \tau e^{i\omega_0 \tau} \beta \Gamma_R\). According to Cauchy argument principle \cite{zhang2020correspondence}, the winding number evaluates to \(w_{\omega_b} = N_z - N_p\) for the reference point \(\omega_b = \omega_0 - i\epsilon\) (\(\epsilon\) is a positive infinitesimal number) located inside \(\mathcal{L}_{\text{BZ}}\) loop, where \(N_z\) and \(N_p\) denote the total number of zeros and poles enclosed within the BZ circle (\(|\beta| = 1\)) for the complex function \(\omega(\beta) - \omega_b\), respectively. For the nearly-unidirectional right-coupling case, we obtain \(N_z = 2\) and \(N_p = 1\), and the winding number is \(w_{\omega_b} = +1\). The detailed calculations can be found in \textbf{SM}.

We further expand the linear-approximation dispersion relation into a polynomial form using the Laurent's expansion:
\begin{equation}
\omega \approx \omega_0 - i\Gamma_t + \frac{N(\beta_1)}{\beta_1 - \beta_2} \sum_{n=0}^{\infty} \beta_1^n \beta^{-(n+1)} + \frac{N(\beta_2)}{\beta_1 - \beta_2} \sum_{n=0}^{\infty} \beta_2^{-(n+1)} \beta^n,
\end{equation}
where \(\beta_1\) and \(\beta_2\) are the roots of \(D(\beta)\) satisfying $|\beta_1\beta_2|=1$, \(|\beta_1|<1\), and \(|\beta_2|>1\). Equation (M11) contains both positive and negative powers of \(\beta\), corresponding to left- and right- coupling channels, respectively. More explicitly, the \(m\)th-order left-right hopping ratio is
\begin{equation}
R_m = \left| \frac{t_m}{t_{-m}} \right| = \left| \frac{N(\beta_2)}{N(\beta_1)} \right| |\beta_1|^2 = \left| \frac{\Gamma_t(\beta_1 - e^{i\omega_0 \tau}) - \Gamma_R \beta_1}{\Gamma_t(\beta_1 - e^{i\omega_0 \tau}) + \Gamma_R e^{i\omega_0 \tau}} \right| ,
\end{equation}
with \(t_m = N(\beta_2)\beta_2^{-(m+1)} / (\beta_1 - \beta_2)\) and \(t_{-m} = N(\beta_1)\beta_1^{m-1} / (\beta_1 - \beta_2)\). Notably, this ratio is independent of the hopping order \(m\), and always satisfies \(R_m > 1\), indicating an effective dominant leftward coupling. This is the origin of the anomalous left-skin localization despite that the non-Markovian system is dominated by the rightward coupling.

\subsection*{Channel-dissipation-induced topological transition}
For the topological transition point, the complex spectrum loop entirely collapses into an infinite line parallel to the real axis. We substitute \(\omega = \omega_r + i\omega_i\) into the non-Markovian nonlinear dispersion relation and extract the imaginary part
\begin{equation}
\omega_i + \Gamma_0 + \frac{\Gamma_L + \Gamma_R}{2} = \Gamma_L + \Gamma_R - \text{Re} \left[ \frac{\Gamma_R}{1 - R_R e^{i(\omega_r \tau_R - q)}} + \frac{\Gamma_L}{1 - R_L e^{i(\omega_r \tau_L + q)}} \right] ,
\end{equation}
with \(R_{L,R} = e^{-(\omega_i + \Gamma_w)\tau_{L,R}}\). Since \(\omega_i\) is independent of \(q\), we have \(\text{Re}[1/(1 - R_{L,R} e^{i(\omega_r \tau_L + q)})] = 1/2\) and \(R_{L,R} = 1\), thus \(\omega_i = -\Gamma_w\). Substituting this back into Eq. (M13), we obtain \(\omega_i = -\Gamma_0\). Consequently, we arrive at the exact topological phase transition condition, \(\Gamma_w = \Gamma_0\), corresponding to the balance between channel loss and site dissipation. At this transition point, the complex spectrum collapses into the line \(\omega = -i\Gamma_0\), with a trivial winding number.

\subsection*{Finite-size non-Bloch theory}
Using the Hamiltonian with retarded long-range coupling in Eq. (M3), we can obtain the second-order recurrence equation,
\begin{equation}
a_{n+2} = \frac{A(1+z_R z_L) - i(\Gamma_R + \Gamma_L)z_R z_L}{z_L(A - i\Gamma_L)} a_{n+1} - \frac{z_R(A - i\Gamma_R)}{z_L(A - i\Gamma_L)} a_n . 
\end{equation}
where \(A = \omega - \widetilde{\omega}_0 + i(\Gamma_L + \Gamma_R)/2\), \(z_R = e^{(i\omega - \Gamma_w)\tau_R}\), \(z_L = e^{(i\omega - \Gamma_w)\tau_L}\). By substituting the ansatz of bulk eigenmodes \(a_n = \beta^n\), we obtain the characteristic equation \(\beta^2 - P\beta + Q = 0\), with \(P = [A(1+z_R z_L) - i(\Gamma_R + \Gamma_L)z_R z_L]/[z_L(A - i\Gamma_L)]\) and \(Q =z_R(A - i\Gamma_R)/[z_L(A - i\Gamma_L)]\). The two characteristic roots are \(\beta_{\pm} = \frac{1}{2} \left[ P \pm \sqrt{P^2 - 4Q} \right]\). 

For the unidirectional right-coupling case (\(\Gamma_L = 0\)), the recurrence relation reduces to the first-order one
\begin{equation}
a_{n+1} = z_R \frac{A - i\Gamma_R}{A} a_n = e^{(i\omega - \Gamma_w)\tau_R} \frac{\omega - \widetilde{\omega}_0 - i\Gamma_R/2}{\omega - \widetilde{\omega}_0 + i\Gamma_R/2} a_n .
\end{equation}
The spatial factor of bulk eigenmodes can be represented by \(\beta = z_R(1 - i\Gamma_R/A)\). Under ABCs, the system behaves as a Jordan chain, and the field accumulates at the rightmost site with \(\beta = \infty\) for \(A=0\) at the \(N\)th-order EP \(\omega_{\mathrm{EP}} = \omega_0 - i\Gamma_t\). Under OBCs, however, the channel reflection provides external leftward feedback with retardation, which satisfies the self-consistent condition \(|\beta|^N |z_L|^N = 1\), giving $\beta = 1/|z_L| = e^{(\omega_i+ \Gamma_w)\tau_L}$. Owing to the hybridization of lattice and channel modes, the system's decay must lie between \(\Gamma_0\) and \(\Gamma_w\). If the memory gain exceeds the channel loss (\(\Gamma_0 > \Gamma_w\)), the retarded feedback field will be stronger, resulting in left-skin localization.

Once a finite left-coupling channel is introduced (\(\Gamma_L \neq 0\)), the recurrence relation recovers the second-order one in Eq. (M14). In this case, the eigenmode is generally a linear superposition of two solutions, \(a_n = c_1 \beta_+^n + c_2 \beta_-^n\). In the limit of \(\Gamma_L \to 0^+\), the first solution continuously approaches the unidirectional right-cascaded mode, \textcolor{black}{\(\beta_+ \to z_R(1 - i\Gamma_R/A)=z_RT_R\) with transmission \(T_R=(1 - i\Gamma_R/A)\),} while the second solution approaches the feedback-dominated spatial branch, \(\beta_- \to 1/z_L\). To illustrate the origin of two branches, we consider the Neumann series expansion of the Green's function \(G(\omega) = (\omega I - H_\text{eff})^{-1} = \sum_{j=0}^{\infty} \frac{1}{A} (W/A)^j\), where $W$ denotes the off-diagonal elements of the effective Hamiltonian in Eq. (M3). For \(\Gamma_L = 0\), the bare right-propogating Green's function \(G_{nm}^R (n > m)\) can be expressed as
\begin{equation}
G_{nm}^R = \frac{1}{A} z_R^{n-m} \sum_{j=1}^{n-m} \binom{n-m-1}{j-1} \left(-\frac{i\Gamma_R}{A}\right)^j = \frac{1}{A} \left(-\frac{i\Gamma_R}{A}\right) z_R^{n-m} \left(1 - \frac{i\Gamma_R}{A}\right)^{n-m-1},
\end{equation}
where the binomial coefficient $\binom{n-m-1}{j-1}$ counts the number of distinct propagation paths involving \(j\) hops between sites \(m\) and \(n\), and \(G_{nn}^R=1/A\). For a small perturbative left coupling \(\Gamma_L\), the first-order Dyson correction can be evaluated using \(G_{nm}^R\) as the unperturbed propagator. At site 1, it is
\begin{equation}
G_{11}^{(1)} = \sum_{l \geq 2} G_{11}^{R} V_{1l} G_{l1}^R = \sum_{l \geq 2} \frac{1}{A} \left(-i\Gamma_L z_L^{l-1}\right) \left(-\frac{i\Gamma_R z_R}{A^2}\right) \beta^{l-2} ,
\end {equation}
with \(V_{1l} = -i\Gamma_L z_L^{l-1}\) and \(\beta = z_R(1 - i\Gamma_R/A)\). Each term \(l\) describes a propagation loop comprising a series of rightward cascadings and one leftward feedback \(1 \to 2 \to \cdots \to l \to 1\): the factor \(-i\Gamma_R z_R \beta^{l-2} / A^2\) represents rightward cascading, \(-i\Gamma_L z_L^{l-1}\) represents the leftward feedback, and \(1/A\) denotes the local response. This process can be visualized in the spatiotemporal lattice of Fig. S4(a), where the purple arrows denote the dressed rightward cascade \(\beta = z_R(1 - i\Gamma_R/A)\), and the pink arrows denote the direct leftward return \(-i\Gamma_L z_L^{l-1}\). In the thermodynamic limit \(N \to \infty\), the first-order Dyson correction becomes \(G_{11}^{(1)}=\frac{\Gamma_L\Gamma_R z_L z_R}{A^3( \beta z_L-1)}\), while the complete Green's function is \(G_{11} = \sum_{j=0}^{\infty} A^{-(j+1)} \Sigma_1^j = \frac{1}{A - \Sigma_1}\), with \(\Sigma_1\) the self-energy. Consequently, we identify the leading self-energy \(\Sigma_1^{(1)} = A^2 G_{11}^{(1)} = \frac{\Gamma_L \Gamma_R z_L z_R}{A(\beta z_L - 1)}\). The eigenvalues are given by the poles of the Green's function, and hence, to leading order in \(\Gamma_L\), satisfy the self-consistent equation \(A-\Sigma_1^{(1)}=0\), namely \(A^2(\beta z_L - 1) = \Gamma_L \Gamma_R z_L z_R\). In the limit \(\Gamma_L \to 0^+\), this condition yields two branches: \(A \to 0\), corresponding to the rightward-cascading Jordan chain at the EP; and \(\beta z_L \to 1\), corresponding to a finite nonzero feedback arising from the coherent interference of all returning loops. 

For a finite system under ABCs, boundary truncation imposes an additional finite-size constraint on the two bulk solutions, leading to an \(N\)-dependent relation between their moduli,
\begin{equation}
\left| \frac{\beta_+}{\beta_-} \right| = \left( \left| \frac{\Gamma_R}{\Gamma_L} \right| \left| \frac{z_R}{z_L} \right| \left| \frac{1 - z_L \beta_+}{z_R - \beta_+} \right|^2 \right)^{1/N} , 
\end{equation}
where \(\left|z_R(1 - z_L \beta_+)/[z_L(z_R - \beta_+)]\right| \sim O(1)\). When \(N \to \infty\), the condition \(|\beta_+| = |\beta_-|\) defines the GBZ, implying that the discrete spectrum under ABCs will gradually converge to the continuous spectrum of the GBZ as the lattice size increases. For the nearly-unidirectional right-coupling system with \(\beta_+ \approx z_R(A - i\Gamma_R)/A\), Eq. (M18) reduces to
\begin{equation}
\left| \frac{\beta_+}{\beta_-} \right| \approx \left[ \frac{\left| 2(1 - z_L z_R)(\omega - \widetilde{\omega}_0) + i \left[ (\Gamma_L + \Gamma_R) + z_L z_R (\Gamma_R - \Gamma_L) \right] \right|^2}{4\Gamma_L \Gamma_R |z_L z_R|} \right]^{1/N} . 
\end{equation}
\textcolor{black}{As \(\omega\) approaches \(\widetilde{\omega}_0\), the modulus separation of two roots is reduced. Since \(|\beta_-| \approx |z_L|^{-1}\) is independent of the detuning \(\Delta = \omega_r - \omega_0\) (up to \(O(\Gamma_L/\Gamma_R)\)), whereas \(|\beta_+| \approx |z_R| |T_R|\) is minimized at \(\Delta = 0\) for \(\Gamma_0 > \Gamma_w\), the reduced modulus separation reflects a suppression of the right-cascaded propagation,} and the spatial profile is primarily determined by the feedback-dominated mode \(|\beta_-| = e^{(\omega_i + \Gamma_w)\tau_L}\). Consequently, when memory gain exceeds the channel loss (\(\Gamma_0 > \Gamma_w\)), \textcolor{black}{the net gain enhances the retarded feedback field}, reconstructing the eigenmode into a rightward-decaying envelope and producing left-skin localization. Conversely, detuning from \(\widetilde{\omega}_0\) makes right-cascaded dynamics dominate, restoring the right-skin localization.

\subsection*{Experiment}
Under the external magnetic bias along the -\textit{y}-direction, the YIG is characterized by a permeability tensor \cite{pozar2011microwave}
\begin{equation}
\mu = \mu_0 \begin{bmatrix}
\mu_r & 0 & -i\kappa_r \\
0 & 1 & 0 \\
i\kappa_r & 0 & \mu_r
\end{bmatrix} ,
\end{equation}
with elements \(\mu_r = 1 + \omega_0 \omega_m / (\omega_0^2 - \omega^2)\) and \(\kappa_r = \omega \omega_m / (\omega_0^2 - \omega^2)\), where Larmor frequency \(\omega_0 = \mu_0 \gamma H_0\) under the external bias \(B_0 = \mu_0 H_0\) with gyromagnetic ratio \(\gamma = 1.579 \times 10^{11}\) C/kg, \(\omega_m = \mu_0 \gamma M_s\) with the saturation magnetization \(4\pi M_s = 1780\) G, and \(\mu_0\) is the vacuum permeability. The relative permittivity of YIG is set as 16.

The system consists of YIG cylinders (\(r = 6.2\) mm, \(h = 11.7\) mm) and a dielectric waveguide (\(420 \times 13 \times 4\) mm\(^3\), \(\varepsilon_r = 12\)) separated by a 2-mm foam spacer, uniformly biased by a 0.1 T static magnetic field along the \(-y\)-direction using paired magnets (diameter: 40 mm, thickness: 5 mm) with 30-mm gap. For the single-cell directional coupling measurements in Fig. 3(b), a far-field horn antenna (VNA Port 1) provides spatial excitation, while the guided signals are collected via impedance-matched WR137 metal waveguides (VNA Ports 2 and 3).

To observe the NHSE [Figs. 4(c, d)], the system is scaled to a 10-cylinder array (period \(d = 42.4\) mm). A meta-cavity antenna replaces the horn to ensure uniform plane-wave illumination with precise incident angle (\(\theta\)) tuning. Copper plates and microwave absorbers are placed at the waveguide terminals to emulate PEC and PML boundaries, respectively. To further demonstrate the loss-induced skin reversal [Fig. 4(e)], we adapt this into a 9-cylinder array with an expanded spacing of 46.4 mm to suppress the site couplings through direct air channels. Crucially, controlled dissipation is introduced by symmetrically attaching 305-mm-long high-loss silicone layers (Laird Eccosorb BSR/MFS; thicknesses 1.28 mm and 1.92 mm) to both sides of the waveguide. This achieves a target effective complex permittivity of \(12 + 0.4i\) at 7.76 GHz, rigorously validated by the excellent agreement between measured (\(S_{11} = -12.31\) dB, \(S_{21} = -24.9\) dB) and simulated (\(S_{11} = -12.12\) dB, \(S_{21} = -20.31\) dB) \(S\)-parameters. Spatial field localization is directly visualized by scanning a \(y\)-polarized near-field probe (VNA Port 2) along the waveguide to record the electric intensity profile (\(|S_{21}|\)) at discrete steps (the details can be found in \textbf{SM}).

\subsection*{Numerical simulations}
All simulations in this work are performed using COMSOL Multiphysics. For the complex band structures of 1D periodic configuration [Figs. 3(c, e)], the eigenfrequency solver was employed on a single unit cell with periodicity along the \(x\)-direction and PML in other directions. This same solver is also used to calculate the eigenfrequencies and eigenstates of the finite-size system under PEC boundaries [Figs. 3(g, h)]. Additionally, driven-response analyses are conducted using the frequency-domain solver. The directionality spectrum [Fig. 3(b)] is derived from \(S\)-parameters via port excitations, and the steady-state field profiles of skin effects [Figs. 4(c–e)] are simulated under background plane-wave illumination.

\subsection*{Author contributions}
S.W. conceived the idea. H.P. performed the numerical simulations and analytical calculations. W.X. and Y.C. assisted in the numerical simulations. K.D. and S.W. assisted in the theory. B.X., Q.L., and Y.W. performed the microwave experiments. H.P., B.X., A.M.H.W., and S.W. analyzed the data. H.P. and S.W. wrote the manuscript with input from all the other authors. S.W. and A.M.H.W. supervised the project. All authors contributed to the discussions.

\subsection*{Funding}
The work described in this paper was supported by grants from  the Research Grants Council of the Hong Kong Special Administrative Region, China (Projects No. 11308223 and No. AoE/P-502/20) and the National Natural Science Foundation of China (No. 12322416). Kun Ding acknowledges funding support from National Key R\&D Program of China (2022YFA1404701, 2022YFA1404500) and the Shanghai Science and Technology Innovation Action Plan (No. 24Z510205936).

\subsection*{Data availability}
The data that support the findings of this study are available from the corresponding authors upon reasonable request.


\clearpage 
\onecolumngrid 

\begin{center}
\textbf{\large Supplementary Material}
\end{center}
\vspace{2ex}

\setcounter{equation}{0}
\setcounter{figure}{0}
\setcounter{table}{0}
\setcounter{page}{1}
\setcounter{section}{0}

\renewcommand{\theequation}{S\arabic{equation}}
\renewcommand{\thefigure}{S\arabic{figure}}
\renewcommand{\thetable}{S\arabic{table}}
\renewcommand{\thesection}{S.\arabic{section}}
\section{Non-Markovian non-Bloch band theory}

We consider a 1D time-translation-invariant lattice with each site containing $D$ internal degrees of freedom, labeled by $\alpha \in \{1, 2, ..., D\}$. Assuming the couplings among different sites are time-retarded, the time-nonlocal equation of motion writes
\begin{equation}
    i\frac{d}{dt}\boldsymbol{a}_n(t) = \sum_{r=-L}^R \int_0^\infty dt' \mathbf{K}_r(t') \boldsymbol{a}_{n+r}(t - t') 
\end{equation}
where $\boldsymbol{a}_n = (a_{n,1}, a_{n,2}, ..., a_{n,D})^T$ with $a_{n,\alpha}^\dagger$ ($a_{n,\alpha}$) denoting the creation (annihilation) operators for the $\alpha$th internal degree of freedom in the $n$th site; $L$ and $R$ specify the maximum coupling ranges to the left and right, respectively;  $\mathbf{K}_r(t')$ is a memory kernel encoding the causal coupling from site $n+r$ at the earlier time $t - t'$ to site $n$ at time $t$. The integration over $t' \ge 0$ guarantees causality and explicitly indicates that the present state depends on the evolution history of other sites. In the Markovian limit of instantaneous coupling, $\mathbf{K}_r(t') = \mathbf{J}_r\delta(t')$, where $\mathbf{J}_r$ is the constant coupling matrix and $\delta(t')$ is the Dirac delta function. For coupling with a constant delay $\tau \ge 0$, corresponding to $\mathbf{K}_r(t') = \mathbf{J}_r\delta(t' - \tau)$, the system is effectively a spatiotemporal lattice connected by coupling channels $(n+r, t-\tau) \to (n, t)$.

By temporal Fourier transformation, the retarded kernel can be converted into the self-energy term, $\mathbf{J}_r(\omega) = \int_0^\infty dt' \mathbf{K}_r(t')e^{i\omega t'}$, and the non-Markovian frequency-dependent Hamiltonian writes
\begin{equation}
    H(\omega) = \sum_{r=-L}^R \mathbf{J}_r(\omega) \boldsymbol{a}_{n+r}^\dagger \boldsymbol{a}_n. 
\end{equation}
Specifically, for the constant delay, $\mathbf{J}_r(\omega) = \mathbf{J}_r e^{i\omega \tau}$, the corresponding Bloch Hamiltonian under periodic boundary condition (PBC) reads $H(q, \omega) = \sum_{r=-L}^R \mathbf{J}_r e^{i(\omega \tau + qr)}$, $q \in [-\pi, \pi)$. The retardation factor $e^{i\omega \tau}$ renders the characteristic equation $F(q, \omega) \equiv \det[\omega\mathbf{I} - H(q, \omega)] = 0$ transcendental, where $\mathbf{I}$ is the identity matrix, thereby generating a multivalued complex dispersion $\omega(q)$ beyond the internal-freedom number $D$. To rigorously characterize these interconnected multivalued branches, we analytically continue the Bloch momentum into the complex plane, $q' = q_r + iq_i$ with $q_r \in [-\pi, \pi)$. Through this extension, the distinct frequency branches constitute multiple sheets of the Riemann surface, defined by the transcendental equation $F(q', \omega) = 0$. Under this geometric framework, the conventional Bloch spectrum for PBC corresponds to the 1D cut along the real momentum axis ($q_i = 0$) of this multi-sheeted manifold. Physically, this multibranch originates from the coupling history encoded in the memory kernel.

The spectrum $\omega(q)$ resides on a multi-sheeted Riemann surface, where a single branch defies conventional Bloch periodicity $\omega(q + 2\pi) = \omega(q)$ and fails to close over a single Brillouin zone (BZ). Consequently, the traditional single-band point-gap winding breaks down. To rigorously capture this emergent topology, the winding number must be defined along a closed spectral orbit $\mathcal{C}$ traversing the Riemann surface. For an orbit that closes after $p$ BZ cycles, parameterized by $s \in [0, 2\pi p)$, the generalized winding number for reference point $\omega_b \notin \mathcal{C}$ is defined as
\begin{equation}
    W_{\mathcal{C}}(\omega_b) = \frac{1}{2\pi i} \int_0^{2\pi p} ds \, \partial_s \log[\omega_{\mathcal{C}}(s) - \omega_b]. 
\end{equation}
This definition can be reduced to the conventional single-band topology ($p = 1$). It provides a rigorous topological measure for the multi-sheeted spectral winding ($p > 1$) intrinsic to non-Markovian dynamics.

Akin to Markovian non-Hermitian systems, for the non-Markovian lattice under open boundary condition (OBC), the conventional Bloch theory breaks down, and a generalized non-Bloch band theory is required to describe its eigen properties. Starting with a single-particle wavefunction $|\psi\rangle = \sum_{n,\alpha} \psi_{n,\alpha}(\omega)\hat{a}_{n,\alpha}^\dagger |0\rangle$ and defining the internal state vector as $\mathbf{\Psi}_n(\omega) = (\psi_{n,1}(\omega), \psi_{n,2}(\omega), \cdots, \psi_{n,D}(\omega))^T$, the frequency-domain equation of motion is cast into $\omega\mathbf{\Psi}_n(\omega) = \sum_{r=-L}^R \mathbf{J}_r(\omega)\mathbf{\Psi}_{n+r}(\omega)$. To properly capture the non-Bloch behavior of the bulk eigenstates, we employ the generalized Bloch ansatz $\mathbf{\Psi}_n = \beta^n \mathbf{u}$. Here, $\beta = e^{iq'}$ introduces a complex momentum $q'$, and $\mathbf{u}$ represents the $D$-component internal degrees of freedom within a unit cell. Consequently, the bulk states are determined by the secular equation $F(\beta, \omega) = 0$.

The characteristic polynomial can be constructed as $P(\beta, \omega) \equiv \beta^{LD} F(\beta, \omega) = 0$, which constitutes an algebraic equation of degree $Q = (L+R)D$ in $\beta$. For a given frequency $\omega$, we sort its $Q$ roots by their moduli in ascending order: $|\beta_1(\omega)| \le |\beta_2(\omega)| \le \cdots \le |\beta_Q(\omega)|$. The eigenstates under OBC are given by the linear superpositions of the generalized Bloch waves, $\psi_n = \sum_{j=1}^Q c_j \beta_j^n \mathbf{u}_j$. Imposing OBC yields $LD$ and $RD$ homogeneous linear constraints at the left and right boundaries, respectively. In the thermodynamic limit $N \to \infty$, the existence of non-trivial solutions requires the asymptotic cancellation of the two most dominant terms in the boundary determinant. This leads to the well-known equal-modulus condition for the continuous spectrum under OBC: $|\beta_{LD}(\omega)| = |\beta_{LD+1}(\omega)|$ \cite{yokomizo2019non}. We note that the nonlinear spectral equation $P(\beta, \omega) = 0$ couples $\omega$ and $\beta$, rendering the roots $\beta_j(\omega)$ multi-valued and causing them to permute across branch cuts. Accordingly, we define the generalized Brillouin zone (GBZ) as
\begin{equation}
    \text{GBZ} := \{\beta \in \mathbb{C}: \exists \omega \in \mathbb{C}, P(\beta, \omega) = 0, |\beta_{LD}(\omega)| = |\beta_{LD+1}(\omega)|\}. 
\end{equation}
The corresponding spectrum $\mathcal{L}_{\text{GBZ}}$ under OBC is the collection of frequencies satisfying the generalized equal-modulus condition:
\begin{equation}
    \mathcal{L}_{\text{GBZ}} := \{\omega \in \mathbb{C}: |\beta_{LD}(\omega)| = |\beta_{LD+1}(\omega)|\}.
\end{equation}
It should be noted that $\mathcal{L}_{\text{GBZ}}$ also manifests as a cut along the GBZ on the multi-sheet Riemann surface defined by $F(q', \omega) = 0$. Thus, this Riemann surface serves as a comprehensive geometric framework, unifying the descriptions of the PBC spectrum, the OBC spectrum, and the underlying multi-branched frequency topology.

\section{Hamiltonian of lattice-continuum coupled system}
The lattice-continuum coupled system can be described by the hybrid Hamiltonian
\begin{equation}
H = \sum_n \tilde{\omega}_0 a_n^\dagger a_n + \frac{1}{2\pi} \int_{-\infty}^{+\infty} dk \omega_k c_k^\dagger c_k + \frac{1}{2\pi} \sum_n \int_{-\infty}^{+\infty} dk (g_k e^{iknd} a_n^\dagger c_k + h.c.).
\end{equation}
where \(a_n^\dagger (a_n)\) and \(c_k^\dagger (c_k)\) are the creation (annihilation) operators for the \(n\)th lattice site and the channel continuum (e.g., waveguide) with wave vector \(k\), and \(d\) is the lattice period. The lattice sites possess a complex on-site potential \(\tilde{\omega}_0 = \omega_0 - i\Gamma_0\) that incorporates intrinsic dissipation, and their coupling to propagation continuum is \(g_k\). By Fourier transformation for the site mode, \(a_n = \frac{1}{\sqrt{N}} \sum_q e^{iqnd} a_q\), with unit-cell number \(N\) and normalized first BZ momentum \(q \in [-\pi, \pi)\), the real-space on-site term in Eq. (S6) can be converted to the momentum-space Hamiltonian as
\begin{equation}
\begin{aligned}
\sum_n \tilde{\omega}_0 a_n^\dagger a_n &= \sum_n \tilde{\omega}_0 \left( \frac{1}{\sqrt{N}} \sum_q e^{-iqn} a_q^\dagger \right) \left( \frac{1}{\sqrt{N}} \sum_{q'} e^{iq'n} a_{q'} \right) \\
&= \frac{\tilde{\omega}_0}{N} \sum_{q, q'} a_q^\dagger a_{q'} \sum_n e^{i(q'-q)n} = \sum_q \tilde{\omega}_0 a_q^\dagger a_q.
\end{aligned} 
\end{equation}
\setcounter{figure}{0} 


To process the continuum channel modes, we discretize the system by considering a finite length \(X = Nd\) and introduce the normalized discrete operators \(C_k = c_k / \sqrt{X}\). Exploiting the discrete translational symmetry of the lattice array with lattice constant \(d\), we adopt the folded-zone representation and decompose the waveguide momentum as \(k = q + mG\). Here, \(q\) lies in the first BZ, \(G = 2\pi\) is the dimensionless reciprocal lattice vector, and \(m\) is the band-folding index. Using the dimensionless index \(q + mG\) to label the folded channel modes, the channel Hamiltonian can be rewritten as
\begin{equation}
 H_{\text{ch}} = \sum_{q,m} \omega_{q+mG} C_{q+mG}^\dagger C_{q+mG}.   
\end{equation}
For the interaction term, by substituting the Fourier expansion of \(a_n\) and applying the lattice summation rule \(\sum_n e^{i(kd-q)n} = N \delta_{kd, q+mG}\) (which enforces quasi-momentum conservation), the coupling term in momentum space is simplified as
\begin{equation}
 H_{\text{int}} = \sum_{q,m} \left( \tilde{g}_{q+mG} a_q^\dagger C_{q+mG} + \tilde{g}_{q+mG}^* C_{q+mG}^\dagger a_q \right),   
\end{equation}
with the rescaled coupling amplitude \(\tilde{g}_{q+mG} = g_{q+mG} / \sqrt{d}\). Therefore, the total Hamiltonian in the folded momentum-space representation is \(H = \sum_q H(q)\), with
\begin{equation}
H(q) = \tilde{\omega}_0 a_q^\dagger a_q + \sum_m \omega_{q+mG} C_{q+mG}^\dagger C_{q+mG} + \sum_m \left( \tilde{g}_{q+mG} a_q^\dagger C_{q+mG} + \tilde{g}_{q+mG}^* C_{q+mG}^\dagger a_q \right). 
\end{equation}
For the unit cell with two internal degrees of freedom, the hybrid Hamiltonian is
\begin{equation}
\begin{aligned}
H &= \sum_n \tilde{\omega}_0^+ a_n^\dagger a_n + \sum_n \tilde{\omega}_0^- b_n^\dagger b_n + \frac{1}{2\pi} \int_{-\infty}^{+\infty} dk \omega_k c_k^\dagger c_k \\
&+ \frac{1}{2\pi} \sum_n \int_{-\infty}^{+\infty} dk (g_{1,k} e^{iknd} a_n^\dagger c_k + g_{1,k}^* e^{-iknd} c_k^\dagger a_n) \\
&+ \frac{1}{2\pi} \sum_n \int_{-\infty}^{+\infty} dk (g_{2,k} e^{iknd} b_n^\dagger c_k + g_{2,k}^* e^{-iknd} c_k^\dagger b_n). 
\end{aligned}
\end{equation}
where \(a_n^\dagger (a_n)\) and \(b_n^\dagger (b_n)\) correspond to different modes in the \(n\)th site with the bare eigenfrequencies \(\tilde{\omega}_0^+\) and \(\tilde{\omega}_0^-\) and respective coupling coefficients to channel modes \(g_{1,k}\) and \(g_{2,k}\). By the spatial Fourier transformation, its corresponding momentum-space form can be obtained as \(H = \sum_q H(q)\), with
\begin{equation}
\begin{aligned}
H(q) &= \tilde{\omega}_0^+ a_q^\dagger a_q + \tilde{\omega}_0^- b_q^\dagger b_q + \sum_m \omega_{q+mG} C_{q+mG}^\dagger C_{q+mG} \\
&+ \sum_m (\tilde{g}_{1,q+mG} a_q^\dagger C_{q+mG} + \tilde{g}_{1,q+mG}^* C_{q+mG}^\dagger a_q) \\
&+ \sum_m (\tilde{g}_{2,q+mG} b_q^\dagger C_{q+mG} + \tilde{g}_{2,q+mG}^* C_{q+mG}^\dagger b_q). 
\end{aligned}
\end{equation}
Because the channel dispersion is folded into the first BZ, the exact block Hamiltonian \(H(q)\) is formally infinite-dimensional: the localized site mode couples to infinite number of folded channel branches labeled by \(m \in \mathbb{Z}\). We project \(H(q)\) in Eq. (S10) onto a finite subspace by imposing a momentum cutoff \(|m| \le M\):
\begin{equation}
\begin{aligned}
H_M(q) &= \tilde{\omega}_0 a_q^\dagger a_q + \sum_{m=-M}^M \omega_{q+mG} C_{q+mG}^\dagger C_{q+mG} \\
&+ \sum_{m=-M}^M (\tilde{g}_{q+mG} a_q^\dagger C_{q+mG} + \tilde{g}_{q+mG}^* C_{q+mG}^\dagger a_q). 
\end{aligned}
\end{equation}
This finite-dimensional approximation is physically justified, as high-order folded modes are far detuned from the emitter resonance frequency, rendering their perturbative contributions to the low-energy modes negligible. The cutoff \(M\) is chosen to ensure convergence of the eigen spectrum within the frequency window of interest. However, we note that this finite basis truncation breaks the translational equivalence at the BZ boundaries, introducing minor discontinuities into the complex band structure at \(q = \pm \pi\), as depicted in Fig. S1.

\renewcommand{\thefigure}{S\arabic{figure}} 
\begin{figure}[htb]
    \centering
    \includegraphics[width=0.6\textwidth]{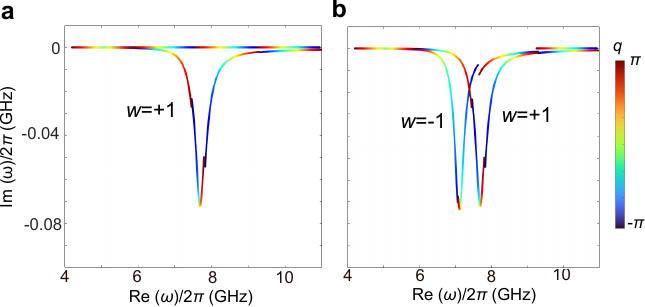}
    \caption{The complex spectra of finite-truncated (\(M=8\)) hybrid Hamiltonian with different lattice internal freedom: (a) single freedom \(\tilde{\omega}_0 = 7.75 - 0.08i\), (b) double freedoms \(\tilde{\omega}_0^+ = 7.15 - 0.08i\) and \(\tilde{\omega}_0^- = 7.75 - 0.08i\). The finite truncation introduces minor discontinuities in the complex band structure at \(q = \pm\pi\).}
\end{figure}

\section{Multipole analysis of cylinder-waveguide coupling}
Here, we derive the explicit expression of the momentum-dependent cylinder-waveguide coupling coefficient, \(g_k\), employing electromagnetic multipole analysis. This is achieved by first expanding the localized cylinder mode into multipoles and then applying angular spectrum representation to project the multipole fields onto the waveguide continuum, following Ref. \cite{vazquez2019near}.

We consider an emitter located at \(\mathbf{r}_0\) in a homogeneous medium (permittivity \(\varepsilon\), permeability \(\mu\)). In the frequency domain with the \(e^{-i\omega t}\) convention, the localized source current can be expressed as \(\mathbf{J}(\mathbf{r}') = -i\omega\mathbf{P}(\mathbf{r}') + \nabla \times \mathbf{M}(\mathbf{r}')\), where \(\mathbf{r}'\) is the local coordinate relative to \(\mathbf{r}_0\). The effective polarization \(\mathbf{P}\) and magnetization \(\mathbf{M}\) distributions are systematically expanded in terms of their respective multipole moments by
\begin{equation}
\mathbf{P}(\mathbf{r}') = \left[\mathbf{p} - \frac{1}{6}\mathbf{Q}^{\text{e}}\nabla + \cdots\right]\delta(\mathbf{r}'), \quad \mathbf{M}(\mathbf{r}') = \left[\mathbf{m} - \frac{1}{6}\mathbf{Q}^{\text{m}}\nabla + \cdots\right]\delta(\mathbf{r}')
\end{equation}
where \(\mathbf{p}\) and \(\mathbf{m}\) are the electric and magnetic dipole moments, and \(\mathbf{Q}^{\text{e}}\) and \(\mathbf{Q}^{\text{m}}\) denote the electric and magnetic quadrupole tensors, respectively.

In the angular spectrum representation, the localized source is decomposed into upward- and downward-propagating (or evanescent) plane-wave components of the form \(e^{i\mathbf{k}^{\pm}\cdot\mathbf{r}}\). Consequently, the spatial differential operator \(\nabla\) acting on each spectral component is equivalently replaced by \(i\mathbf{k}^{\pm}\), where \(\mathbf{k}^{\pm} = (k_x, k_y, \pm k_z)\) denotes the wave vector for propagation into the \(z > z_0\) (+) and \(z < z_0\) (-) regions. By applying this substitution, the total source current associated with different multipoles can be elegantly recast into the form of an effective electric-dipole current, \(\mathbf{J}(\mathbf{r}') = -i\omega\mathbf{p}_{\text{eff}}\delta(\mathbf{r}')\). The explicit forms of the effective electric dipole moments \(\mathbf{p}_{\text{eff}}^{(\alpha)}\) for the electric dipole (ED) \(\mathbf{p}\), magnetic dipole (MD) \(\mathbf{m}\), electric quadrupole (EQ) \(\mathbf{Q}^{\text{e}}\), and magnetic quadrupole (MQ) \(\mathbf{Q}^{\text{m}}\) naturally emerge as
\begin{equation}
\begin{aligned}
\mathbf{p}_{\text{eff}}^{(\text{ED})} &= \mathbf{p}, & \mathbf{p}_{\text{eff}}^{(\text{EQ})} &= -\frac{i}{6}\mathbf{Q}^{\text{e}} \cdot \mathbf{k}^{\pm}, \\
\mathbf{p}_{\text{eff}}^{(\text{MD})} &= -\frac{1}{\omega}\mathbf{k}^{\pm} \times \mathbf{m}, & \mathbf{p}_{\text{eff}}^{(\text{MQ})} &= \frac{i}{6\omega}\mathbf{k}^{\pm} \times (\mathbf{Q}^{\text{m}} \cdot \mathbf{k}^{\pm}).
\end{aligned}
\end{equation}

The longitudinal wave-vector component is dictated by the dispersion relation, \(k_z = \sqrt{k_0^2 - k_t^2}\) for propagating modes (\(k_t \le k_0\)) and \(k_z = i\sqrt{k_t^2 - k_0^2}\) for evanescent modes (\(k_t > k_0\)), with \(k_0 = \omega\sqrt{\varepsilon\mu}\) and \(k_t = \sqrt{k_x^2 + k_y^2}\). By utilizing these effective moments, the angular spectra of the radiated electric and magnetic fields evaluated at the source plane \(z = z_0\) can be compactly expressed in a universal form as
\begin{equation}
\begin{aligned}
\mathbf{E}(k_x, k_y)|_{z=z_0} &= \frac{ik_0^2}{8\pi^2\varepsilon k_z} \sum_{\alpha} \left[ \left(\mathbf{\hat{e}}_s \cdot \mathbf{p}_{\text{eff}}^{(\alpha)}\right)\mathbf{\hat{e}}_s + \left(\mathbf{\hat{e}}_p^{\pm} \cdot \mathbf{p}_{\text{eff}}^{(\alpha)}\right)\mathbf{\hat{e}}_p^{\pm} \right], \\
\mathbf{H}(k_x, k_y)|_{z=z_0} &= \frac{i\omega k_0}{8\pi^2 k_z} \sum_{\alpha} \left[ \left(\mathbf{\hat{e}}_p^{\pm} \cdot \mathbf{p}_{\text{eff}}^{(\alpha)}\right)\mathbf{\hat{e}}_s - \left(\mathbf{\hat{e}}_s \cdot \mathbf{p}_{\text{eff}}^{(\alpha)}\right)\mathbf{\hat{e}}_p^{\pm} \right].
\end{aligned}
\end{equation}
where the unit vectors for \(s\)-polarized (TE) and \(p\)-polarized (TM) waves are defined as \(\mathbf{\hat{e}}_s = (-k_y/k_t, k_x/k_t, 0)\) and \(\mathbf{\hat{e}}_p = (\pm k_x k_z/(k_0 k_t), \pm k_y k_z/(k_0 k_t), -k_t/k_0)\). This unified formulation provides a straightforward avenue to determine the coupling strengths. The fraction of source power coupled into a specific waveguide mode is directly proportional to the projection of the corresponding effective dipole moment onto the modal polarization. Consequently, the excited amplitudes for TE and TM waveguide modes scale with \(|\mathbf{\hat{e}}_s \cdot \mathbf{p}_{\text{eff}}^{(\alpha)}|\) and \(|\mathbf{\hat{e}}_p^{\pm} \cdot \mathbf{p}_{\text{eff}}^{(\alpha)}|\), respectively.
\begin{figure}
    \centering
    \includegraphics[width=0.9\textwidth]{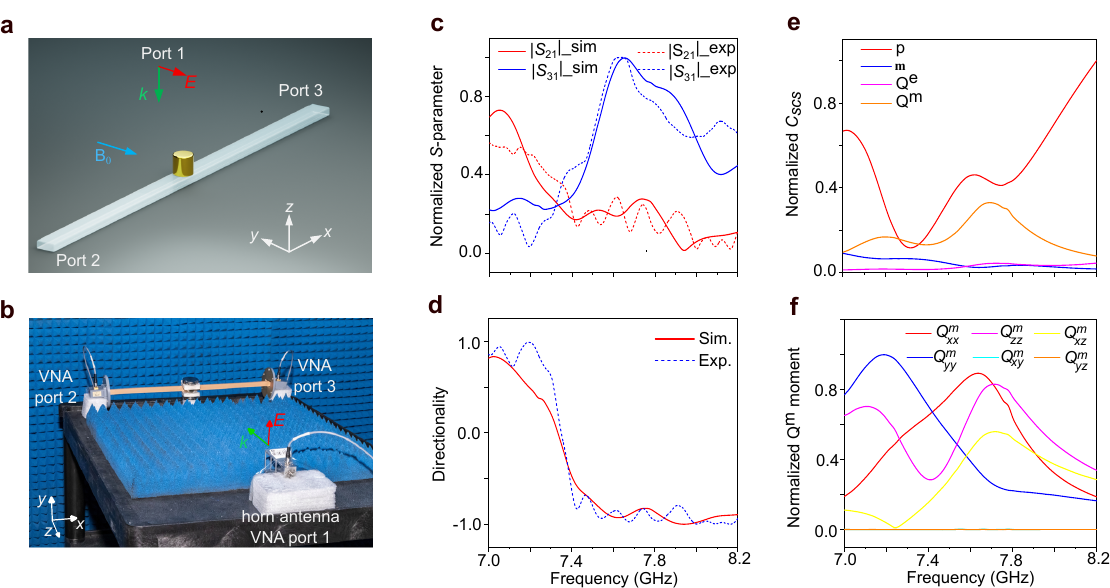}
    \caption{The chiral-coupling characteristic of single YIG cylinder-waveguide system with the external magnetic bias. (a) Schematic of the 1D single YIG cylinder-waveguide coupling system, which enables asymmetric and nonreciprocal conversion between free-space plane waves and TE guided modes under an external magnetic bias. (b) Experimental setup configuration for directionality measurement. The pair of magnets provides the uniform external magnetic bias of \(B_0=0.1\) T along -\(y\)-direction. (c) Simulated and experimental scattering parameters \(|S_{21}|\) and \(|S_{31}|\) versus frequency for the YIG-cylinder-waveguide structure excited by a linearly polarized free-space wave with the electric field along the \(y\)-direction. (d) Corresponding directionality spectrum defined as \((|S_{21}|^2-|S_{31}|^2)/(|S_{21}|^2+|S_{31}|^2)\). (e) Simulated scattering cross section \(C_{\text{scs}}\) of different multipole moments, where \(\mathbf{p}\), \(\mathbf{m}\), \(\mathbf{Q}^{\text{e}}\), and \(\mathbf{Q}^{\text{m}}\) are electric dipole, magnetic dipole, electric quadrupole, and magnetic quadrupole moments, respectively. (f) Simulated magnetic quadrupole tensor elements versus frequency.}
\end{figure}
Importantly, in the evanescent regime (\(k_t > k_0\)) typical for guided modes, \(k_z\) becomes purely imaginary. This renders the polarization vector \(\mathbf{\hat{e}}_p^{\pm}\) complex, effectively representing an elliptical or circular polarization in momentum space. Under such conditions, specific multipole configurations, such as a circularly polarized dipole \(\mathbf{p} = (1, 0, \pm i)\), yield highly asymmetric or even nearly unidirectional coupling. These analytical results seamlessly recover the chiral light-matter interaction behaviors derived from Fermi's golden rule \cite{picardi2018janus}, while naturally extending the predictive capability to higher-order multipolar interfaces. Especially, for the circular magnetic quadrupole moments located above the waveguide, \(\mathbf{Q}_{\pm}^{\text{m}} = (1, 0, \pm i; 0, 0, 0; \pm i, 0, 1)\), their coupling strength to 1D TE waveguide mode with \(\mathbf{k} = (k_m, 0, i\alpha_m)\) based on Eq. (S16) can be represented by \(g_{k_m} \propto |(\alpha_m \pm k_m)|^2/k_0^2\) with \(\alpha_m = \sqrt{k_m^2 - k_0^2}\).

The YIG material under the magnetic-field bias along -\(y\)-direction possesses the permeability tensor \cite{pozar2011microwave}, \(\mu = \mu_0[\mu_r, 0, -i\kappa_r; 0, 1, 0; i\kappa_r, 0, \mu_r]\), which breaks the \(x\)-direction mirror symmetry, thus inducing the directional coupling to waveguide propagating modes. For the microwave single cylinder-waveguide coupled system with the external bias \(B_0=0.1\) T depicted in Fig. S2(a), the linear-polarized plane wave with \(E_y\) component can be directionally converted to the TE guided mode, as shown by the scattering parameters \(S_{21}\) and \(S_{31}\) in Fig. S2(c). The numerically simulated directionality, defined as \((|S_{21}|^2-|S_{31}|^2)/(|S_{21}|^2+|S_{31}|^2)\), changes from nearly +1 to -1 when the frequency increases, as shown in Fig. S2(d), signifying a reversal of the coupling directionality. This directional switching is corroborated by the experimental measurement [Fig. S2(b)], which exhibits excellent agreement with the numerical simulations. To unravel the physical genesis of this directional coupling, Fig. S2(e) shows the normalized multipole scattering cross sections (\(C_{\text{scs}}\)) of the YIG cylinder. While the resonance is co-dominated by electric dipole and magnetic quadrupole modes, angular spectrum analysis reveals that the excited \(y\)-oriented electric dipole (\(p_y\)) symmetrically couples with the TE guided mode of the waveguide (\(k_y=0\)). Its inherent elliptical magnetic quadrupole moments (i.e., \(Q_{zz}^{\text{m}}/Q_{xx}^{\text{m}} = 3.04e^{i54.23^\circ}\), \(Q_{xz}^{\text{m}}/Q_{xx}^{\text{m}} = 0.477e^{i51.6^\circ}\) at 7.06 GHz and \(Q_{zz}^{\text{m}}/Q_{xx}^{\text{m}} = 0.9e^{i159.9^\circ}\), \(Q_{xz}^{\text{m}}/Q_{xx}^{\text{m}} = 0.6e^{-i112.55^\circ}\) at 7.66 GHz), presented in Fig. S2(f), induce an asymmetric excitation of the TE guided modes propagating in opposite directions, which is a direct manifestation of the spin-momentum locking effect.

\section{Non-Markovian effective model with long-range coupling}

For the 1D lattice coupled to a continuum channel described by Eq. (S6), we can integrate out the channel's degrees of freedom to obtain an effective lattice-subspace non-Hermitian Hamiltonian that encodes the channel-mediated interactions between the localized sites.

Starting from the Heisenberg equations of motion for the lattice mode \(a_n\) and the channel continuum mode \(c_k\) (setting \(\hbar=1\)), i.e., \(idO/dt = [O, H]\), the coupled dynamics can be described by
\begin{equation}
\begin{aligned}
i\frac{da_n(t)}{dt} &= \tilde{\omega}_0 a_n(t) + \frac{1}{2\pi}\int_{-\infty}^{+\infty} d k g_k e^{iknd} c_k(t), \\
i\frac{dc_k(t)}{dt} &= \omega_k c_k(t) + \sum_m g_k^* e^{-ikmd} a_m(t).
\end{aligned}
\end{equation}
We transform the equations to frequency domain through temporal Fourier transformation:
\begin{equation}
\begin{aligned}
(\omega - \tilde{\omega}_0)a_n(\omega) &= \frac{1}{2\pi}\int_{-\infty}^{+\infty} d k g_k e^{iknd} c_k(\omega),\\
(\omega - \omega_k)c_k(\omega) &= \sum_m g_k^* e^{-ikmd} a_m(\omega).
\end{aligned}
\end{equation}
To ensure causality, we apply \(\omega \rightarrow \omega + i0^+\) to shift the pole below the real axis and obtain
\begin{equation}
c_k(\omega) = \sum_m \frac{g_k^* e^{-ikmd}}{\omega - \omega_k + i0^+} a_m(\omega).
\end{equation}
Substituting this back into the frequency-domain equation of \(a_n(\omega)\), we obtain
\begin{equation}
(\omega - \tilde{\omega}_0)a_n(\omega) = \sum_m J_{nm}(\omega) a_m(\omega),
\end{equation}
where the channel-mediated coupling term is \(J_{nm}(\omega) = \frac{1}{2\pi}\int_{-\infty}^{+\infty} d k \frac{|g_k|^2 e^{ikd(n-m)}}{\omega - \omega_k + i0^+}\). The dynamics are thus governed by a frequency-dependent effective non-Hermitian Hamiltonian
\begin{equation}
H_{\text{eff}}(\omega) = \sum_n \tilde{\omega}_0 a_n^\dagger a_n + \sum_{n,m} J_{nm}(\omega) a_n^\dagger a_m.
\end{equation}
We assume a linear dispersion of the continuum channel
\begin{equation}
\omega_k = \begin{cases} v_g^R k, & k > 0 \\ -v_g^L k, & k < 0 \end{cases},
\end{equation}
with right and left group velocity \(v_g^R\) and \(v_g^L\), and constant couplings between emitters and propagation channel
\begin{equation}
g_k = \begin{cases} g_R, & k > 0, \\ g_L, & k < 0, \end{cases}
\end{equation}
where \(g_R\) and \(g_L\) denote the coupling strengths to right- and left-propagating modes, respectively. The couplings among different sites can be expressed as
\begin{equation}
J_{nm}(\omega) = -\frac{1}{2\pi v_g} \left[ |g_R|^2 \int_0^{+\infty} d k \frac{e^{ikd(n-m)}}{k - k_p^R + i0^+} + |g_L|^2 \int_0^{+\infty} d k' \frac{e^{-ik'd(n-m)}}{k' - k_p^L + i0^+} \right],
\end{equation}
with \(k_p^R = \omega/v_g^R\) and \(k_p^L = \omega/v_g^L\). The integrals can be evaluated by contour
integration in the complex \(k\)-plane. The pole contribution gives the radiative, direction-dependent coupling between emitters. The principal-value part gives a coherent frequency renormalization. For \(n\neq m\), with the
linear dispersion and frequency-independent coupling, the nonlocal principal-value contribution vanishes after extending the momentum integration to \((-\infty,+\infty)\). For \(n=m\), it reduces to a local,
channel-induced Lamb shift, which is absorbed into the definition of the bare emitter frequency. In what follows we omit this Lamb-shift term and retain only the pole contribution, thus acquiring
\begin{equation}
J_{nm}(\omega) = -i\Gamma_R e^{i\omega(n-m)\tau_R} \Theta(n-m) - i\Gamma_L e^{i\omega(m-n)\tau_L} \Theta(m-n),
\end{equation}
where \(\Gamma_R = |g_R|^2/v_g^R\) and \(\Gamma_L = |g_L|^2/v_g^L\) denote the right and left coupling via the channel, respectively, and \(\tau_R = d/v_g^R\) and \(\tau_L = d/v_g^L\) represent the unit right and left retardation time, and \(\Theta(x)\) is the Heaviside step function with \(\Theta(0)=1/2\). Therefore, Eq. (S21) becomes
\begin{equation}
\begin{aligned}
H_{\text{eff}}(\omega) = \sum_n \left(\tilde{\omega}_0 - i\frac{\Gamma_L+\Gamma_R}{2}\right) a_n^\dagger a_n 
- i\Gamma_R \sum_{n>m} e^{i\omega(n-m)\tau_R} a_n^\dagger a_m - i\Gamma_L \sum_{n<m} e^{i\omega(m-n)\tau_L} a_n^\dagger a_m.
\end{aligned}
\end{equation}
The frequency-dependent phase factors in the effective Hamiltonian signify an inherently retarded channel-mediated interaction, a direct consequence of the finite propagation velocity of the channel mode. The inverse temporal Fourier transform of these factors to the time domain maps them onto delayed amplitudes, \(\frac{1}{2\pi}\int_{-\infty}^{+\infty} d\omega a_m(\omega) e^{i\omega\tau} e^{-i\omega t} = a_m(t-\tau)\), thus yielding
\begin{equation}
i\frac{da_n(t)}{dt} = \left(\tilde{\omega}_0 - i\frac{\Gamma_L+\Gamma_R}{2}\right) a_n(t) - i\Gamma_R \sum_{n>m} a_m(t-\tau_{nm}^R) - i\Gamma_L \sum_{n<m} a_m(t-\tau_{nm}^L),
\end{equation}
with right and left retarded time \(\tau_{nm}^R = |n-m|\tau_R\) and \(\tau_{nm}^L = |n-m|\tau_L\), respectively, which explicitly capture the nonlocal-in-time nature of the coupling: the finite propagation delay dictates that the current state of any given site depends on the past states of the other sites, rendering the effective dynamics fundamentally non-Markovian.

In the limit that the system's characteristic evolution timescale far exceeds the propagation delay, these memory effects become negligible. By invoking the Markovian approximation \(a_m(t-\tau_{nm}^{R/L}) \simeq e^{i\omega_0\tau_{nm}^{R/L}}a_m(t)\), the temporal retardation results in a pure spatial phase shift. Consequently, the system simplifies into a Markovian case characterized by instantaneous, phase-modulated long-range couplings, described by the effective Hamiltonian
\begin{equation}
H_{\text{eff}}^{\text{M}} = \sum_n \left(\tilde{\omega}_0 - i\frac{\Gamma_L+\Gamma_R}{2}\right) a_n^\dagger a_n - i\Gamma_R \sum_{n>m} e^{i\omega_0\tau_{nm}^R} a_n^\dagger a_m - i\Gamma_L \sum_{n<m} e^{i\omega_0\tau_{nm}^L} a_n^\dagger a_m.
\end{equation}
If dissipation is introduced into the coupling channel, its dispersion relation becomes
\begin{equation}
\omega_k = \begin{cases} v_g^R k - i\Gamma_w, & k > 0 \\ -v_g^L k - i\Gamma_w, & k < 0 \end{cases},
\end{equation}
with dissipation rate \(\Gamma_w\), and the effective non-Markovian Hamiltonian in Eq. (S26) should be modified as
\begin{equation}
\begin{aligned}
H_{\text{eff}}(\omega) &= \sum_n \left(\tilde{\omega}_0 - i\frac{\Gamma_L+\Gamma_R}{2}\right) a_n^\dagger a_n \\
&- i\Gamma_R \sum_{n>m} e^{(i\omega-\Gamma_w)\tau_{nm}^R} a_n^\dagger a_m - i\Gamma_L \sum_{n<m} e^{(i\omega-\Gamma_w)\tau_{nm}^L} a_n^\dagger a_m,
\end{aligned}
\end{equation}
Meanwhile, the Markovian-limit Hamiltonian in Eq. (S28) becomes
\begin{equation}
\begin{aligned}
H_{\text{eff}}^{\text{M}}(\omega) &= \sum_n \left(\tilde{\omega}_0 - i\frac{\Gamma_L+\Gamma_R}{2}\right) a_n^\dagger a_n \\
&- i\Gamma_R \sum_{n>m} e^{(i\omega_0-\Gamma_w)\tau_{nm}^R} a_n^\dagger a_m - i\Gamma_L \sum_{n<m} e^{(i\omega_0-\Gamma_w)\tau_{nm}^L} a_n^\dagger a_m.
\end{aligned}
\end{equation}
Furthermore, by mapping the on-site and coupling terms of the Hamiltonian in Eq. (S30) into the reciprocal space, the Hamiltonian can be represented with independent Bloch modes, yielding \(H_{\text{eff}}(\omega) = \sum_q H(q, \omega) a_q^\dagger a_q\), with the momentum-space Hamiltonian
\begin{equation}
H(q, \omega) = \tilde{\omega}_0 - i\frac{\Gamma_L+\Gamma_R}{2} - i\Gamma_R \sum_{l=1}^{\infty} e^{(i\omega-\Gamma_w)l\tau_R} e^{-ilq} - i\Gamma_L \sum_{l=1}^{\infty} e^{(i\omega-\Gamma_w)l\tau_L} e^{ilq},
\end{equation}
where \(l\) denotes the distance between two coupled sites in units of the lattice constant. 
The infinite-range nature of the channel-mediated interactions leads to geometric series in momentum space. 
To make the analytic structure explicit, we define \(z_R(\omega,q)=e^{(i\omega-\Gamma_w)\tau_R}e^{-iq}\) and \(
z_L(\omega,q)=e^{(i\omega-\Gamma_w)\tau_L}e^{iq}.
\)
In the domain where \(|z_R|<1\) and \(|z_L|<1\), these series converge to
\(\sum_{l=1}^{\infty}z_R^l=\frac{z_R}{1-z_R}\) and \(\sum_{l=1}^{\infty}z_L^l=\frac{z_L}{1-z_L}.
\)
The closed-form expressions on the right-hand side can be understood as the causal meromorphic continuation of the retarded waveguide self-energy. 
This continuation is fixed by the retarded boundary condition, or equivalently by the \(+i0^+\) prescription of the underlying waveguide Green's function, and introduces no additional approximation. 
Therefore, the self-consistent dispersion relation is
\begin{equation}
\omega
=
\widetilde{\omega}_0
-i\frac{\Gamma_L+\Gamma_R}{2}
-i\Gamma_R
\frac{e^{(i\omega-\Gamma_w)\tau_R}e^{-iq}}
{1-e^{(i\omega-\Gamma_w)\tau_R}e^{-iq}}
-i\Gamma_L
\frac{e^{(i\omega-\Gamma_w)\tau_L}e^{iq}}
{1-e^{(i\omega-\Gamma_w)\tau_L}e^{iq}} .
\end{equation}
The corresponding momentum-space Markovian Hamiltonian is \(H_{\text{eff}}^{\text{M}}(\omega) = \sum_q H^{\text{M}}(q, \omega) a_q^\dagger a_q\) with
\begin{equation}
H^{\text{M}}(q, \omega) = \tilde{\omega}_0 - i\frac{\Gamma_L+\Gamma_R}{2} - i\Gamma_R \sum_{l=1}^{\infty} e^{(i\omega_0-\Gamma_w)l\tau_R} e^{-ilq} - i\Gamma_L \sum_{l=1}^{\infty} e^{(i\omega_0-\Gamma_w)l\tau_L} e^{ilq}.
\end{equation}
Therefore, the Markovian dispersion relation can be expressed as
\begin{equation}
\omega = \tilde{\omega}_0 - i\frac{\Gamma_L+\Gamma_R}{2} - i\Gamma_R \frac{e^{(i\omega_0-\Gamma_w)\tau_R}e^{-iq}}{1-e^{(i\omega_0-\Gamma_w)\tau_R}e^{-iq}} - i\Gamma_L \frac{e^{(i\omega_0-\Gamma_w)\tau_L}e^{iq}}{1-e^{(i\omega_0-\Gamma_w)\tau_L}e^{iq}}.
\end{equation}
When \(\Gamma_w \rightarrow 0\), it is simplified as
\begin{equation}
\omega(q) = \tilde{\omega}_0 + \frac{\Gamma_R}{2}\cot((q_R-q)/2) + \frac{\Gamma_L}{2}\cot((q_L+q)/2),
\end{equation}
with \(q_R = \omega_0\tau_R\) and \(q_L = \omega_0\tau_L\). Crucially, the imaginary parts of the eigenfrequencies are pinned to a constant, \(\text{Im}(\omega(q)) = -\Gamma_0\). As a result, the complex energy spectrum collapses into an open, infinite, straight line parallel to the real axis. This highlights a fundamental physical constraint in the Markovian system with infinite-range coupling: the instantaneous infinite-range coupling leads to a topological singularity with closed point gap. The waveguide dissipation \(\Gamma_w\) regulates this singularity and enables the complex spectral to form loops with nontrivial winding numbers \cite{yu2022giant}.

Similarly, for the case with each site containing two internal degrees of freedom, the effective Hamiltonian can be expressed as
\begin{equation}
\begin{aligned}
H_{\text{eff}}(\omega) &= \sum_n \left(\tilde{\omega}_0^+ - i\frac{\Gamma_L^+ + \Gamma_R^+}{2}\right) a_n^\dagger a_n - i\Gamma_R^+ \sum_{n>m} e^{(i\omega-\Gamma_w)\tau_{nm}^R} a_n^\dagger a_m - i\Gamma_L^+ \sum_{n<m} e^{(i\omega-\Gamma_w)\tau_{nm}^L} a_n^\dagger a_m \\
&+ \sum_n \left(\tilde{\omega}_0^- - i\frac{\Gamma_L^- + \Gamma_R^-}{2}\right) b_n^\dagger b_n - i\Gamma_R^- \sum_{n>m} e^{(i\omega-\Gamma_w)\tau_{nm}^R} b_n^\dagger b_m - i\Gamma_L^- \sum_{n<m} e^{(i\omega-\Gamma_w)\tau_{nm}^L} b_n^\dagger b_m \\
&- i\frac{\Lambda_L^{+-} + \Lambda_R^{+-}}{2} \sum_n a_n^\dagger b_n - i\Lambda_R^{+-} \sum_{n>m} e^{(i\omega-\Gamma_w)\tau_{nm}^R} a_n^\dagger b_m - i\Lambda_L^{+-} \sum_{n<m} e^{(i\omega-\Gamma_w)\tau_{nm}^L} a_n^\dagger b_m \\
&- i\frac{\Lambda_L^{-+} + \Lambda_R^{-+}}{2} \sum_n b_n^\dagger a_n - i\Lambda_R^{-+} \sum_{n>m} e^{(i\omega-\Gamma_w)\tau_{nm}^R} b_n^\dagger a_m - i\Lambda_L^{-+} \sum_{n<m} e^{(i\omega-\Gamma_w)\tau_{nm}^L} b_n^\dagger a_m,
\end{aligned}
\end{equation}
with
\begin{equation}
\begin{aligned}
\Gamma_L^+ &= \frac{|g_L^+|^2}{v_g}, \Gamma_R^+ = \frac{|g_R^+|^2}{v_g}, \Gamma_L^- = \frac{|g_L^-|^2}{v_g}, \Gamma_R^- = \frac{|g_R^-|^2}{v_g}, \\
\Lambda_L^{+-} &= \frac{g_L^+ g_L^{-*}}{v_g}, \Lambda_R^{+-} = \frac{g_R^+ g_R^{-*}}{v_g}, \Lambda_L^{-+} = \frac{g_L^- g_L^{+*}}{v_g}, \Lambda_R^{-+} = \frac{g_R^- g_R^{+*}}{v_g}.
\end{aligned}
\end{equation}
where \(g_L^i\) and \(g_R^i\) (\(i=+,-\)) denote the coupling coefficients between internal freedom \(i\) in the emitters and left and right propagating channels, respectively. Based on the Fourier transformation, the corresponding momentum-space Hamiltonian can be expressed as
\begin{equation}
H_{\text{eff}}(\omega) = \sum_q (a_q^\dagger, b_q^\dagger) H(q) \begin{pmatrix} a_q \\ b_q \end{pmatrix}, \quad H(q) = \begin{pmatrix} H_{aa} & H_{ab} \\ H_{ba} & H_{bb} \end{pmatrix},
\end{equation}
with
\begin{equation}
\begin{aligned}
H_{aa}(q) &= \tilde{\omega}_0^+ - i\frac{\Gamma_L^+ + \Gamma_R^+}{2} - i\Gamma_R^+ \frac{e^{(i\omega-\Gamma_w)\tau_R}e^{-iq}}{1-e^{(i\omega-\Gamma_w)\tau_R}e^{-iq}} - i\Gamma_L^+ \frac{e^{(i\omega-\Gamma_w)\tau_L}e^{iq}}{1-e^{(i\omega-\Gamma_w)\tau_L}e^{iq}}, \\
H_{ab}(q) &= -i\frac{\Lambda_L^{+-} + \Lambda_R^{+-}}{2} - i\Lambda_R^{+-} \frac{e^{(i\omega-\Gamma_w)\tau_R}e^{-iq}}{1-e^{(i\omega-\Gamma_w)\tau_R}e^{-iq}} - i\Lambda_L^{+-} \frac{e^{(i\omega-\Gamma_w)\tau_L}e^{iq}}{1-e^{(i\omega-\Gamma_w)\tau_L}e^{iq}}, \\
H_{ba}(q) &= -i\frac{\Lambda_L^{-+} + \Lambda_R^{-+}}{2} - i\Lambda_R^{-+} \frac{e^{(i\omega-\Gamma_w)\tau_R}e^{-iq}}{1-e^{(i\omega-\Gamma_w)\tau_R}e^{-iq}} - i\Lambda_L^{-+} \frac{e^{(i\omega-\Gamma_w)\tau_L}e^{iq}}{1-e^{(i\omega-\Gamma_w)\tau_L}e^{iq}}, \\
H_{bb}(q) &= \tilde{\omega}_0^- - i\frac{\Gamma_L^- + \Gamma_R^-}{2} - i\Gamma_R^- \frac{e^{(i\omega-\Gamma_w)\tau_R}e^{-iq}}{1-e^{(i\omega-\Gamma_w)\tau_R}e^{-iq}} - i\Gamma_L^- \frac{e^{(i\omega-\Gamma_w)\tau_L}e^{iq}}{1-e^{(i\omega-\Gamma_w)\tau_L}e^{iq}}.
\end{aligned}
\end{equation}

\section{Effective Hamiltonian of finite-size system}

For the finite-size (\(N\)) lattice under absorbing boundary conditions (ABCs), the effective Hamiltonian is
\begin{equation}
\begin{aligned}
H_{\text{eff}}^{\text{ABC}} = &\sum_{n=1}^N \left( \tilde{\omega}_0 - i\frac{\Gamma_L + \Gamma_R}{2} \right) a_n^\dagger a_n - i\Gamma_R \sum_{1 \le m < n \le N} e^{(i\omega - \Gamma_w)(n-m)\tau_R} a_n^\dagger a_m \\
&- i\Gamma_L \sum_{1 \le n < m \le N} e^{(i\omega - \Gamma_w)(m-n)\tau_L} a_n^\dagger a_m.
\end{aligned}
\end{equation}
The ABCs allow the coupling signal to propagate without reflection at the ends, mimicking the infinite continuum channel, which is distinct from the OBCs. A rigorous realization of the non-Markovian OBC requires the truncations of both the lattice and the continuum channel. Consequently, the effective inter-site coupling must take into account the multiple reflections of propagation waves.

To evaluate the coupling elements \( J_{nm}^{\text{OBC}}(\omega) \) for \( m < n \), we sum over all propagation waves between these sites. The leftmost and the rightmost sites are labeled as 0 and \(N+1\). The direct right-propagating wave and its round trips yield a geometric series
\begin{equation}
-i\Gamma_R e^{(i\omega - \Gamma_w)(n-m)\tau_R} \sum_{p=0}^\infty e^{(i\omega - \Gamma_w)(N+1)p(\tau_R +\tau_L)} = \frac{-i\Gamma_R e^{(i\omega - \Gamma_w)(n-m)\tau_R} }{1 - e^{(i\omega - \Gamma_w)(N+1)(\tau_R + \tau_L)}},
\end{equation}
where $p$ denotes the $p$th round trip. The left-propagating wave with even-number reflections also induces a geometric series, \( -i\Gamma_L e^{(i\omega - \Gamma_w)(m\tau_L + (N+1)\tau_R + (N+1-n)\tau_L)} / (1 - e^{(i\omega - \Gamma_w)(N+1)(\tau_R + \tau_L)}) \). Furthermore, waves involving an odd number of reflections result in the geometric series, \( i\Lambda e^{(i\omega - \Gamma_w)(m\tau_L + n\tau_R)} / (1 - e^{(i\omega - \Gamma_w)(N+1)(\tau_R + \tau_L)}) \) and \( i\Lambda^* e^{(i\omega - \Gamma_w)((N+1-m)\tau_R + (N+1-n)\tau_L)} / (1 - e^{(i\omega - \Gamma_w)(N+1)(\tau_R + \tau_L)}) \), with the cross-coupling amplitudes \( \Lambda = g_L^* g_R / \sqrt{v_g^R v_g^L} \) and \( \Lambda^* = g_R^* g_L / \sqrt{v_g^R v_g^L} \). Summing over these four distinct classes of waves, we obtain the total coupling
\begin{equation}
\begin{aligned}
J_{nm}^{\text{OBC}}(\omega) = -i \big( &\Gamma_R e^{(i\omega - \Gamma_w)(n-m)\tau_R} + \Gamma_L e^{(i\omega - \Gamma_w)[m\tau_L + (N+1)\tau_R + (N+1-n)\tau_L]} - \Lambda e^{(i\omega - \Gamma_w)(m\tau_L + n\tau_R)} \\
&- \Lambda^* e^{(i\omega - \Gamma_w)[(N+1-m)\tau_R + (N+1-n)\tau_L]} \big) / (1 - e^{(i\omega - \Gamma_w)(N+1)(\tau_R + \tau_L)}).
\end{aligned}
\end{equation}
The corresponding coupling term for \( m > n \) is obtained similarly by interchanging the left and right propagating channels:
\begin{equation}
\begin{aligned}
J_{nm}^{\text{OBC}}(\omega) = -i \big( &\Gamma_R e^{(i\omega - \Gamma_w)[(N+1-m+n)\tau_R + (N+1)\tau_L]} + \Gamma_L e^{(i\omega - \Gamma_w)(m-n)\tau_L} - \Lambda e^{(i\omega - \Gamma_w)(n\tau_R + m\tau_L)} \\
&- \Lambda^* e^{(i\omega - \Gamma_w)[(N+1-m)\tau_R + (N+1-n)\tau_L]} \big) / (1 - e^{(i\omega - \Gamma_w)(N+1)(\tau_R + \tau_L)}).
\end{aligned}
\end{equation}
Finally, the on-site term (\( m = n \)) comprises the local radiative decay into both channels, i.e., \( -\frac{i}{2}(\Gamma_R + \Gamma_L) \), which is dressed by all complete round trips between the boundaries, as well as the odd-reflection feedback paths involving the cross-coupling terms \( \Lambda \) and \( \Lambda^* \):
\begin{equation}
\begin{aligned}
J_{nn}^{\text{OBC}}(\omega) = -i \Big[ &\frac{\Gamma_R + \Gamma_L}{2} (1 + e^{(i\omega - \Gamma_w)(N+1)(\tau_R + \tau_L)}) - \Lambda e^{(i\omega - \Gamma_w)n(\tau_R + \tau_L)} \\
&- \Lambda^* e^{(i\omega - \Gamma_w)(N+1-n)(\tau_R + \tau_L)} \Big] / (1 - e^{(i\omega - \Gamma_w)(N+1)(\tau_R + \tau_L)})
\end{aligned}
\end{equation}
The effective Hamiltonian under OBCs is
\begin{equation}
H_{\text{eff}}^{\text{OBC}}(\omega) = \sum_{n=1}^N \tilde{\omega}_0 a_n^\dagger a_n + \sum_{1 \le n, m \le N} J_{nm}^{\text{OBC}}(\omega) a_n^\dagger a_m,
\end{equation}
with
\begin{equation}
\begin{aligned}
J_{nm}^{\text{OBC}}(\omega) &= -\frac{i}{1 - e^{(i\omega - \Gamma_w)(N+1)(\tau_R + \tau_L)}} \times \\
&\quad \begin{cases}
\begin{aligned}
&\Gamma_R e^{(i\omega - \Gamma_w)(n-m)\tau_R} + \Gamma_L e^{(i\omega - \Gamma_w)[m\tau_L + (N+1)\tau_R + (N+1-n)\tau_L]} \\
&- \Lambda e^{(i\omega - \Gamma_w)(m\tau_L + n\tau_R)} - \Lambda^* e^{(i\omega - \Gamma_w)[(N+1-m)\tau_R + (N+1-n)\tau_L]},
\end{aligned} & m < n \\
\begin{aligned}
&\frac{\Gamma_R + \Gamma_L}{2} (1 + e^{(i\omega - \Gamma_w)(N+1)(\tau_R + \tau_L)}) \\
&- \Lambda e^{(i\omega - \Gamma_w)n(\tau_R + \tau_L)} - \Lambda^* e^{(i\omega - \Gamma_w)(N+1-n)(\tau_R + \tau_L)},
\end{aligned} & m = n \\
\begin{aligned}
&\Gamma_L e^{(i\omega - \Gamma_w)(m-n)\tau_L} + \Gamma_R e^{(i\omega - \Gamma_w)[(N+1-m+n)\tau_R + (N+1)\tau_L]} \\
&- \Lambda e^{(i\omega - \Gamma_w)(m\tau_L + n\tau_R)} - \Lambda^* e^{(i\omega - \Gamma_w)[(N+1-m)\tau_R + (N+1-n)\tau_L]},
\end{aligned} & m > n
\end{cases}
\end{aligned}
\end{equation}

Similarly, for the case with two internal degrees of freedom per unit cell, the effective Hamiltonian under ABCs can be expressed as
\begin{equation}
\begin{aligned}
H_{\text{eff}}^{\text{ABC}} &= 
\sum_{n=1}^N \left( \tilde{\omega}_0^+ - i\frac{\Gamma_L^+ + \Gamma_R^+}{2} \right) a_n^\dagger a_n \\
&\quad - i\Gamma_R^+ \sum_{1 \le m < n \le N} e^{(i\omega - \Gamma_w)(n-m)\tau_R} a_n^\dagger a_m - i\Gamma_L^+ \sum_{1 \le n < m \le N} e^{(i\omega - \Gamma_w)(m-n)\tau_L} a_n^\dagger a_m \\
&\quad + \sum_{n=1}^N \left( \tilde{\omega}_0^- - i\frac{\Gamma_L^- + \Gamma_R^-}{2} \right) b_n^\dagger b_n \\
&\quad - i\Gamma_R^- \sum_{1 \le m < n \le N} e^{(i\omega - \Gamma_w)(n-m)\tau_R} b_n^\dagger b_m - i\Gamma_L^- \sum_{1 \le n < m \le N} e^{(i\omega - \Gamma_w)(m-n)\tau_L} b_n^\dagger b_m \\
&\quad - i\frac{\Lambda_L^{+-} + \Lambda_R^{+-}}{2} \sum_{n=1}^N a_n^\dagger b_n \\
&\quad - i\Lambda_R^{+-} \sum_{1 \le m < n \le N} e^{(i\omega - \Gamma_w)(n-m)\tau_R} a_n^\dagger b_m - i\Lambda_L^{+-} \sum_{1 \le n < m \le N} e^{(i\omega - \Gamma_w)(m-n)\tau_L} a_n^\dagger b_m \\
&\quad - i\frac{\Lambda_L^{-+} + \Lambda_R^{-+}}{2} \sum_{n=1}^N b_n^\dagger a_n \\
&\quad - i\Lambda_R^{-+} \sum_{1 \le m < n \le N} e^{(i\omega - \Gamma_w)(n-m)\tau_R} b_n^\dagger a_m - i\Lambda_L^{-+} \sum_{1 \le n < m \le N} e^{(i\omega - \Gamma_w)(m-n)\tau_L} b_n^\dagger a_m,
\end{aligned}
\end{equation}
and the OBC effective Hamiltonian can be written as
\begin{equation}
H_{\text{eff}}^{\text{OBC}} = \sum_{1 \le n, m \le N} (a_n^\dagger \quad b_n^\dagger) \begin{pmatrix} H_{nm}^{aa} & H_{nm}^{ab} \\ H_{nm}^{ba} & H_{nm}^{bb} \end{pmatrix} \begin{pmatrix} a_m \\ b_m \end{pmatrix},
\end{equation}
where \( \alpha, \beta = a, b \) denotes different degrees of freedom in a unit cell, and
\begin{equation}
H_{nm}^{\alpha\beta} = \delta_{\alpha, \beta} \delta_{n, m} \tilde{\omega}_0^\alpha + J_{nm}^{\alpha\beta},
\end{equation}
with
\begin{equation}
\begin{aligned}
J_{nm}^{\alpha\beta} &= -\frac{i}{1 - e^{(i\omega - \Gamma_w)(N+1)(\tau_R + \tau_L)}} \times \\
&\quad \begin{cases}
\begin{aligned}
&\Gamma_R^{\alpha\beta} e^{(i\omega - \Gamma_w)(n-m)\tau_R} + \Gamma_L^{\alpha\beta} e^{(i\omega - \Gamma_w)[m\tau_L + (N+1)\tau_R + (N+1-n)\tau_L]} \\
&- \Lambda_{RL}^{\alpha\beta} e^{(i\omega - \Gamma_w)(m\tau_L + n\tau_R)} - \Lambda_{LR}^{\alpha\beta} e^{(i\omega - \Gamma_w)[(N+1-m)\tau_R + (N+1-n)\tau_L]},
\end{aligned} & m < n \\
\begin{aligned}
&\frac{\Gamma_R^{\alpha\beta} + \Gamma_L^{\alpha\beta}}{2} (1 + e^{(i\omega - \Gamma_w)(N+1)(\tau_R + \tau_L)}) \\
&- \Lambda_{RL}^{\alpha\beta} e^{(i\omega - \Gamma_w)n(\tau_R + \tau_L)} - \Lambda_{LR}^{\alpha\beta} e^{(i\omega - \Gamma_w)(N+1-n)(\tau_R + \tau_L)},
\end{aligned} & m = n \\
\begin{aligned}
&\Gamma_L^{\alpha\beta} e^{(i\omega - \Gamma_w)(m-n)\tau_L} + \Gamma_R^{\alpha\beta} e^{(i\omega - \Gamma_w)[(N+1-m+n)\tau_R + (N+1)\tau_L]} \\
&- \Lambda_{RL}^{\alpha\beta} e^{(i\omega - \Gamma_w)(m\tau_L + n\tau_R)} - \Lambda_{LR}^{\alpha\beta} e^{(i\omega - \Gamma_w)[(N+1-m)\tau_R + (N+1-n)\tau_L]},
\end{aligned} & m > n
\end{cases}
\end{aligned}
\end{equation}
and \( \Gamma_R^{\alpha\beta} = g_R^\alpha g_R^{\beta*} / v_g^R \), \( \Gamma_L^{\alpha\beta} = g_L^\alpha g_L^{\beta*} / v_g^L \), \( \Lambda_{LR}^{\alpha\beta} = g_L^\alpha g_R^{\beta*} / \sqrt{v_g^R v_g^L} \), and \( \Lambda_{RL}^{\alpha\beta} = g_R^\alpha g_L^{\beta*} / \sqrt{v_g^R v_g^L} \).

\section{Linear-approximation effective model}

As detailed in S4, the retardation effects in the non-Markovian long-range coupling model manifest as frequency-dependent couplings, yielding a nonlinear eigenvalue problem. Mapping this model to a Markovian picture via linear approximation allows us to understand its non-Hermitian spectral topology within a conventional framework.

For the case that each emitter contains one internal degree of freedom and there is no dissipation in the continuum channel, the nonlinear dispersion can be simplified as \(\omega - \omega_0 = -i\Gamma_t + f(\omega, \beta)\) with the total on-site dissipation \(\Gamma_t = \Gamma_0 + (\Gamma_L + \Gamma_R)/2\) and coupling term \(f(\omega, \beta) = -i\Gamma_R e^{i\omega\tau_R}/(\beta - e^{i\omega\tau_R}) - i\Gamma_L e^{i\omega\tau_L}\beta/(1 - e^{i\omega\tau_L}\beta)\), assuming the Bloch phase factor \(\beta = e^{iq}\). Applying Taylor expansion  to the coupling term \(f(\omega, \beta)\) at the reference frequency \(\omega_0\), we obtain
\begin{equation}
f(\omega, \beta) = f(\omega_0, \beta) + (\omega - \omega_0)\frac{\partial f}{\partial \omega}\Big|_{\omega_0} + \frac{1}{2}(\omega - \omega_0)^2\frac{\partial^2 f}{\partial \omega^2}\Big|_{\omega_0} + \cdots,
\end{equation}
where \(\partial f/\partial \omega|_{\omega_0} = f_1(\omega_0, \beta)\) and \(\partial^2 f/\partial \omega^2|_{\omega_0} = f_2(\omega_0, \beta)\) with

\begin{equation}
\begin{aligned}
f_1(\omega_0, \beta) &= \beta \left[ \frac{\Gamma_R\tau_R e^{i\omega_0\tau_R}}{(\beta - e^{i\omega_0\tau_R})^2} + \frac{\Gamma_L\tau_L e^{i\omega_0\tau_L}}{(1 - e^{i\omega_0\tau_L}\beta)^2} \right], \\
f_2(\omega_0, \beta) &= i\Gamma_R\tau_R^2\beta e^{i\omega_0\tau_R} \left( \frac{2e^{i\omega_0\tau_R}}{(\beta - e^{i\omega_0\tau_R})^3} + \frac{1}{(\beta - e^{i\omega_0\tau_R})^2} \right) \\
&\quad + i\Gamma_L\tau_L^2\beta e^{i\omega_0\tau_L} \left( \frac{2e^{i\omega_0\tau_L}\beta}{(1 - e^{i\omega_0\tau_L}\beta)^3} + \frac{1}{(1 - e^{i\omega_0\tau_L}\beta)^2} \right).
\end{aligned}
\end{equation}
We retain up to the first order in Eq. (S52), which can faithfully capture the topological features under the convergence condition \(|(\omega - \omega_0)\tau_{L,R}| < 1\), provided that the higher-order retardation terms do not close the relevant point gap. The resulting linearized dispersion relation is
\begin{equation}
\omega - \omega_0 \approx \frac{-i\Gamma_t + f(\omega_0, \beta)}{1 - f_1(\omega_0, \beta)} = \frac{-i\Gamma_t - i\Gamma_R \frac{e^{i\omega_0\tau_R}}{\beta - e^{i\omega_0\tau_R}} - i\Gamma_L \frac{e^{i\omega_0\tau_L}\beta}{1 - e^{i\omega_0\tau_L}\beta}}{1 - \beta \left[ \frac{\Gamma_R\tau_R e^{i\omega_0\tau_R}}{(\beta - e^{i\omega_0\tau_R})^2} + \frac{\Gamma_L\tau_L e^{i\omega_0\tau_L}}{(1 - e^{i\omega_0\tau_L}\beta)^2} \right] }  .
\end{equation}
When \(\tau_{L,R} \to 0\), the linearized dispersion reduces to the Markovian dispersion in Eq. (S36) while preserving the phase factor. Notably, this linear approximation is equivalent to the single Newton-iteration with the initial value \(\omega = \omega_0\), i.e., \(\omega_{\text{new}} = \omega_0 - g(\omega_0, \beta)/g'(\omega_0, \beta) = \omega_0 + (-i\Gamma_t + f(\omega_0, \beta))/(1 - f_1(\omega_0, \beta))\) with \(g(\omega, \beta) = \omega_0 - i\Gamma_t + f(\omega, \beta) - \omega\).

For the nearly unidirectional right-coupling case (\(\Gamma_R \gg \Gamma_L\)) with symmetric retardation \(\tau_L = \tau_R = \tau\), the linearized dispersion can be simplified as \(\omega - \omega_0 \approx (-i\Gamma_t(\beta - e^{i\omega_0\tau})^2 - i\Gamma_R e^{i\omega_0\tau}(\beta - e^{i\omega_0\tau})) / ((\beta - e^{i\omega_0\tau})^2 - \tau e^{i\omega_0\tau}\beta\Gamma_R)\). To analyze the point-gap topology, we consider the winding of the complex band around a reference point \(\omega_b\), whose winding number is defined as \(w_{\omega_b} = \frac{1}{2\pi i} \oint_{|\beta|=1} d\beta \frac{\partial}{\partial \beta} \log[\omega(\beta) - \omega_b]\). To ensure \(\omega_b\) lies within the complex spectral loop, we set \(\omega_b = \omega_0 - i\epsilon\) with \(\epsilon \to 0^+\). According to the Cauchy argument principle, the winding number of any complex function evaluated along the unit circle \(|\beta| = 1\) is given by \(w_{\omega_b} = N_z - N_p\), where \(N_z\) and \(N_p\) denote the number of zeros and poles enclosed within the unit circle \cite{zhang2020correspondence}. Therefore, we rewrite the complex function as a rational function of \(\beta\), i.e.,
\begin{equation}
\omega - \omega_b = i\epsilon + \frac{N(\beta)}{D(\beta)} = \frac{N(\beta) + i\epsilon D(\beta)}{D(\beta)},
\end{equation}
with numerator \(N(\beta) = -i\Gamma_t(\beta - e^{i\omega_0\tau})^2 - i\Gamma_R e^{i\omega_0\tau}(\beta - e^{i\omega_0\tau})\) and denominator \(D(\beta) = (\beta - e^{i\omega_0\tau})^2 - \tau e^{i\omega_0\tau}\beta\Gamma_R\). The pole number \(N_p\) is determined by the roots of \(D(\beta)\), which, according to Vieta's formulas, satisfy \(|\beta_1^p \beta_2^p| = |e^{i2\omega_0\tau}| = 1\). However, since \(\tau\Gamma_R > 0\), the two roots cannot simultaneously reside on the unit circle, and one pole must lie inside the circle, yielding \(N_p = 1\). Meanwhile, the number of zeros \(N_z\) is determined by \(N(\beta) + i\epsilon D(\beta)\). In the unperturbed limit (\(\epsilon = 0\)), the roots of \(N(\beta)\) can be expressed as \(\beta_1^z = (\Gamma_t - \Gamma_R)e^{i\omega_0\tau}/\Gamma_t\) and \(\beta_2^z = e^{i\omega_0\tau}\), with the moduli \(|\beta_1^{z}| = |(\Gamma_0 - \Gamma_R/2)/(\Gamma_0 + \Gamma_R/2)| < 1\) and \(|\beta_2^{z}| = |e^{i\omega_0\tau}| = 1\). While a weak perturbation \(\epsilon\) shifts the root \(\beta_1^z\) inside the unit circle, and \(\beta_2^{(z)}\) lies exactly on the unit circle at \(\epsilon = 0\), necessitating an explicit evaluation of its perturbative displacement. Therefore, we introduce the ansatz \(\beta_2^{(z)} = e^{i\omega_0\tau} + \delta\), with the perturbation-induced shift \(\delta\). Substituting this into the zero-point condition and expanding to the leading order, we get the asymptotic equation:
\begin{equation}
N(\beta_2^z) + i\epsilon D(\beta_2^z) = -i\Gamma_R e^{i\omega_0\tau}\delta - i\epsilon \tau e^{2i\omega_0\tau}\Gamma_R = 0,
\end{equation}
From which we obtain \(\delta = -\epsilon\tau e^{i\omega_0\tau}\), which gives the shifted root \(\beta_2^{(z)} = (1 - \epsilon\tau)e^{i\omega_0\tau}\). For \(\epsilon\tau > 0\), this zero point is pushed inside the unit circle. Consequently, the weak perturbation ensures that both zeros are enclosed by the unit circle, leading to \(N_z = 2\). Thus, the winding number evaluates to \(w_{\omega_b} = N_z - N_p = 1\), consistent with the numerical spectra in Fig. S3(b). This result is consistent with the non-Markovian effective model, demonstrating that the linear approximation successfully captures the system's underlying topological features. Similarly, for the case of nearly unidirectional left-coupling model (\(\Gamma_R \ll \Gamma_L\)), the calculated zero and pole numbers are 0 and 1, respectively, and thus \(w_{\omega_b} = N_z - N_p = -1\), which agrees with the numerical result in Fig. S3(a).
\begin{figure}
    \centering
   \includegraphics[width=0.6\textwidth]{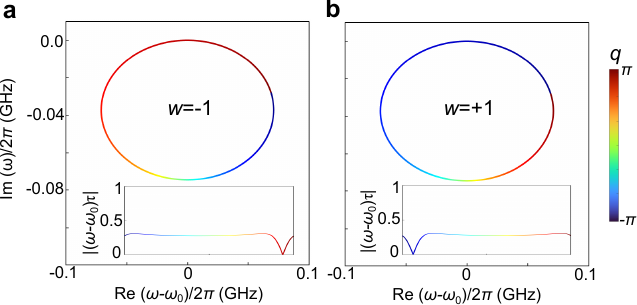}
    \caption{The linear-approximation complex spectrum for nearly unidirectional (a) left (\(\Gamma_R \ll \Gamma_L\)) and (b) right (\(\Gamma_R \gg \Gamma_L\)) coupling cases. The insets show \(|(\omega - \omega_0)\tau|\) that satisfies the convergent condition.}
\end{figure}

Moreover, we can expand the linearized dispersion into a polynomial form to intuitively understand the mechanism of the anomalous non-Hermitian topological phase. According to the roots of \(D(\beta)\) (\(|\beta_1^p| < 1\) and \(|\beta_2^p| > 1\)), the linearized dispersion can be decomposed into \(\omega \approx \omega_0 - i\Gamma_t + C_1/(\beta - \beta_1) + C_2/(\beta - \beta_2)\) with \(C_1 = N(\beta_1)/(\beta_1 - \beta_2)\) and \(C_2 = N(\beta_2)/(\beta_2 - \beta_1)\), assuming \(\beta_1 = \beta_1^p\) and \(\beta_2 = \beta_2^p\) for simplicity. Due to their distinct pole locations relative to the unit circle, these two fractional terms necessitate different Laurent expansions, i.e., \(C_1/(\beta - \beta_1) = C_1 \sum_{n=0}^\infty \beta_1^n \beta^{-(n+1)}\) and \(C_2/(\beta - \beta_2) = -C_2 \sum_{n=0}^\infty \beta_2^{-(n+1)} \beta^n\), indicating that the inner (\(\beta_1\)) and outer (\(\beta_2\)) poles drive the right and left couplings, denoted by the negative and positive powers of \(\beta\), respectively. More explicitly, the \(m\)th-order left-right hopping ratio is
\begin{equation}
R_m = \left| \frac{t_m}{t_{-m}} \right| = \left| \frac{N(\beta_2)}{N(\beta_1)} \right| |\beta_1|^2 = \left| \frac{\Gamma_t(\beta_1 - e^{i\omega_0\tau}) - \Gamma_R\beta_1}{\Gamma_t(\beta_1 - e^{i\omega_0\tau}) + \Gamma_R e^{i\omega_0\tau}} \right|,
\end{equation}
with \(t_m = N(\beta_2)\beta_2^{-(m+1)}/(\beta_1 - \beta_2)\) and \(t_{-m} = N(\beta_1)\beta_1^{m-1}/(\beta_1 - \beta_2)\). Notably, this ratio is independent of the hopping order \(m\). In the passive regime (\(\Gamma_0 > 0\)), it yields \(R_m > 1\), corresponding to an effective dominant left coupling. Conversely, it yields \(R_m < 1\) for the nearly-unidirectional left-coupling scenario. This reveals the origin of the anomalous topological phase, demonstrating that the interference of different memory layers inherent in retarded long-range couplings can effectively reverse the dominant coupling direction within the Markovian picture.

\section{Finite-size non-Bloch theory}
For the effective Hamiltonian with retarded long-range couplings, the corresponding
eigenvalue problem \(H_{\rm eff}(\omega)|\psi\rangle=\omega|\psi\rangle\) can be
projected onto the real-space lattice basis. Writing the eigenstate as
\(|\psi\rangle=\sum_n a_n a_n^\dagger|0\rangle\), one obtains
\begin{equation}
A a_n
=
-i\Gamma_R \sum_{m<n} z_R^{n-m} a_m
-i\Gamma_L \sum_{m>n} z_L^{m-n} a_m ,
\end{equation}
where
\(
A=\omega-\tilde{\omega}_0+i (\Gamma_L + \Gamma_R)/2\), 
\(z_R=e^{(i\omega-\Gamma_w)\tau_R}\), and 
\(z_L=e^{(i\omega-\Gamma_w)\tau_L}\).
To convert this non-local real-space equation into a local recurrence relation,
we introduce two auxiliary fields
\(
R_n=\sum_{m<n}z_R^{n-m}a_m\) and
\(L_n=\sum_{m>n}z_L^{m-n}a_m .
\)
Equation~(S58) can be rewritten as
\(
A a_n+i\Gamma_R R_n+i\Gamma_L L_n=0.
\)
Introducing the shift operator \(T\), defined by \((Ta)_n=a_{n+1}\), these
relations can be compactly written as
\((T-z_R)R=z_R a\) and \((1-z_L T)L=z_L T a\).
Acting on the eigenvalue equation with \((T-z_R)(1-z_L T)\), one obtains a purely
local second-order recurrence relation for \(a_n\):
\begin{equation}
z_L(i\Gamma_L-A)a_{n+2}
+\left[A(1+z_R z_L)-i(\Gamma_R+\Gamma_L)z_R z_L\right]a_{n+1}
+z_R(i\Gamma_R-A)a_n=0 .
\end{equation}
Provided that \(z_L(i\Gamma_L-A)\neq 0\), Eq.~(S59) can be rewritten as
\begin{equation}
a_{n+2}
=
\frac{A(1+z_R z_L)-i(\Gamma_R+\Gamma_L)z_R z_L}
{z_L(A-i\Gamma_L)}a_{n+1}
-
\frac{z_R(A-i\Gamma_R)}
{z_L(A-i\Gamma_L)}a_n .
\end{equation}
This recurrence equation determines the spatial distribution of the bulk eigenmodes.
Substituting the ansatz \(a_n=\beta^n\) into the recurrence equation, we obtain the characteristic equation
\(\beta^2-P\beta+Q=0\), with
\[
P=
\frac{A(1+z_R z_L)-i(\Gamma_R+\Gamma_L)z_R z_L}
{z_L(A-i\Gamma_L)},
\qquad
Q=
\frac{z_R(A-i\Gamma_R)}
{z_L(A-i\Gamma_L)} .
\]
The two characteristic roots are
\(
\beta_{\pm}
=\frac{1}{2}
\left[
P\pm\sqrt{P^2-4Q}
\right]
\),
or explicitly,
\begin{equation}
\begin{aligned}
\beta_{\pm} = \frac{1}{2} \Bigg[ & \frac{A(1+z_R z_L) - i(\Gamma_R + \Gamma_L)z_R z_L}{z_L(A - i\Gamma_L)} \\
& \quad \pm \sqrt{ \left( \frac{A(1+z_R z_L) - i(\Gamma_R + \Gamma_L)z_R z_L}{z_L(A - i\Gamma_L)} \right)^2 - 4 \frac{z_R(A - i\Gamma_R)}{z_L(A - i\Gamma_L)} } \Bigg]
\end{aligned}
\end{equation}

For the unidirectional right-coupling case \((\Gamma_L=0)\), the original
non-local equation reduces to
\(A a_n=-i\Gamma_R R_n\),
together with
\((T-z_R)R=z_R a\).
The system is then governed by a first-order recurrence relation,
\begin{equation}
a_{n+1}
=
z_R\frac{A-i\Gamma_R}{A}a_n
=
e^{(i\omega-\Gamma_w)\tau_R}
\frac{\omega-\tilde{\omega}_0-i\Gamma_R/2}
{\omega-\tilde{\omega}_0+i\Gamma_R/2}
a_n .
\end{equation}
Therefore, for \(a_n=\beta^n\), the spatial factor is
\(\beta=z_R(A-i\Gamma_R)/A\).
Under the ABCs, the field finally accumulates at the rightmost site, \(\beta=z_R(A-i\Gamma_R)/A=\infty\),
with \(A=0\) at the \(N\)th-order EP point
\(
\omega_{\rm EP}=\omega_0-i\Gamma_t\).
Under the OBCs, however, the truncation of the continuum channel generates an external
left feedback with the retardation factor \(z_L\), which should satisfy the self-consistent condition
\(|\beta|^N |z_L|^N=1\), giving \(|\beta|=1/|z_L|=e^{(\omega_i+\Gamma_w)\tau_L}\).
Owing to hybridization of the site and continuum modes, the system's decay must lie 
between \(\Gamma_0\) and \(\Gamma_w\). If the memory gain dominates the
channel loss \((\Gamma_0>\Gamma_w)\), the retarded feedback field will be
amplified, thus exhibiting left-skin localization.

Once a finite left-coupling is introduced, namely \(\Gamma_L\neq 0\),
the system dynamics 
recovers the second-order recurrence relation, and the spatial profile of eigenmodes is generally a superposition of two modes,
\(
a_n=c_1\beta_+^n+c_2\beta_-^n
\).
In the nearly unidirectional limit \(\Gamma_L\to 0\), we have 
\(\beta_+\to z_R(A-i\Gamma_R)/A\) and \(\beta_-\to 1/z_L\).
The first root \(\beta_+\) continuously connects to the unidirectional right-cascaded mode with transmission \(T_R=(A-i\Gamma_R)/A\), while the second root \(\beta_-\) corresponds to the
leftward feedback channel. Physically, \(\beta_-\) originates from the coherent superposition of multiple retarded feedback fields. To illustrate the origin of two branches, we consider the Neumann series expansion of the Green’s function \(G(\omega) = (\omega - H_{eff})^{-1} = \sum_{j=0}^{\infty} \frac{1}{A} (W/A)^j\), where \(W\) denotes the off-diagonal elements of the effective Hamiltonian in Eq. (S30). For \(\Gamma_L = 0\), the combination of right-hopping term can induce the bare right Green's function \(G_{nm}^R (n > m)\):
\begin{equation}
G_{nm}^R = \frac{1}{A} z_R^{n-m} \sum_{j=1}^{n-m} \binom{n-m-1}{j-1} \left(-\frac{i\Gamma_R}{A}\right)^j = \frac{1}{A} \left(-\frac{i\Gamma_R}{A}\right) z_R^{n-m} \left(1 - \frac{i\Gamma_R}{A}\right)^{n-m-1},
\end{equation}
where the binomial coefficient $\binom{n-m-1}{j-1}$ counts the number of distinct propagation paths involving \(j\) hops between sites \(m\) and \(n\), and \(G_{nn}^R=1/A\). By treating the left coupling \(\Gamma_L\) as a perturbation, the first-order term in the Dyson series expansion of the Green's function can be evaluated using \(G_{nm}^R\) as the unperturbed (zeroth-order) propagator. Specifically, the first-order correction to the local Green's function at site 1, \(G_{11}^{(1)}\), is given by
\begin{equation}
G_{11}^{(1)} = \sum_{l \geq 2} G_{11}^R V_{1l} G_{l1}^R = \sum_{l \geq 2} \frac{1}{A} (-i\Gamma_L z_L^{l-1}) \left(-\frac{i\Gamma_R z_R}{A^2}\right) \beta^{l-2} ,
\end{equation}
with the perturbed left-coupling term \(V_{1l} = -i\Gamma_L z_L^{l-1}\) and \(\beta = z_R(1 - i\Gamma_R/A)\). Each term \(l\) in the sum is one right-cascade-left-feedback loop \(1 \to 2 \to \cdots \to l \to 1\): \(-i\Gamma_R z_R \beta^{l-2}/A^2\) describes the right-cascaded process, \(-i\Gamma_L z_L^{l-1}\) denotes the left feedback, and \(1/A\) is the local-site response, which can be visualized on the spatiotemporal lattice of Fig. S4(a). In the time domain, \(z_{R,L} = e^{(i\omega - \Gamma_w)\tau_{R,L}}\) act as retarded propagators, \(z_R^{n-m} a_m \to e^{-\Gamma_w(n-m)\tau_R} a_m(t - (n-m)\tau_R)\), so the loop closing at site \(l\) carries the total round-trip delay \((l-1)(\tau_R + \tau_L)\). The purple arrows denote the dressed rightward cascade \(\beta = z_R(1 - i\Gamma_R/A)\), while the pink arrows denote the direct leftward return \(-i\Gamma_L z_L^{l-1}\). In the thermodynamic limit \(N \to \infty\), the first-order Dyson correction becomes \(G_{11}^{(1)}=\frac{\Gamma_L\Gamma_R z_L z_R}{A^3( \beta z_L-1)}\), while the complete Green's function is \(G_{11} = \sum_{j=0}^{\infty} A^{-(j+1)} \Sigma_1^j = \frac{1}{A - \Sigma_1}\), with \(\Sigma_1\) the self-energy. Consequently, we identify the leading self-energy \(\Sigma_1^{(1)} = A^2 G_{11}^{(1)} = \frac{\Gamma_L \Gamma_R z_L z_R}{A(\beta z_L - 1)}\). The eigenvalues are given by the poles of the Green's function, and hence, to leading order in \(\Gamma_L\), satisfy the self-consistent equation \(A-\Sigma_1^{(1)}=0\), namely \(A^2(\beta z_L - 1) = \Gamma_L \Gamma_R z_L z_R\). When \(\Gamma_L \to 0^+\), it has two distinct solutions: the \(A \to 0\) denoting the EP characterized by right-localized Jordan chains, and the \(\beta z_L \to 1\) branch where an infinitesimal \(\Gamma_L\) sustains finite feedback via the coherent divergence of all returning loops.

\begin{figure}[htb]
    \centering
    \includegraphics[width=\linewidth]{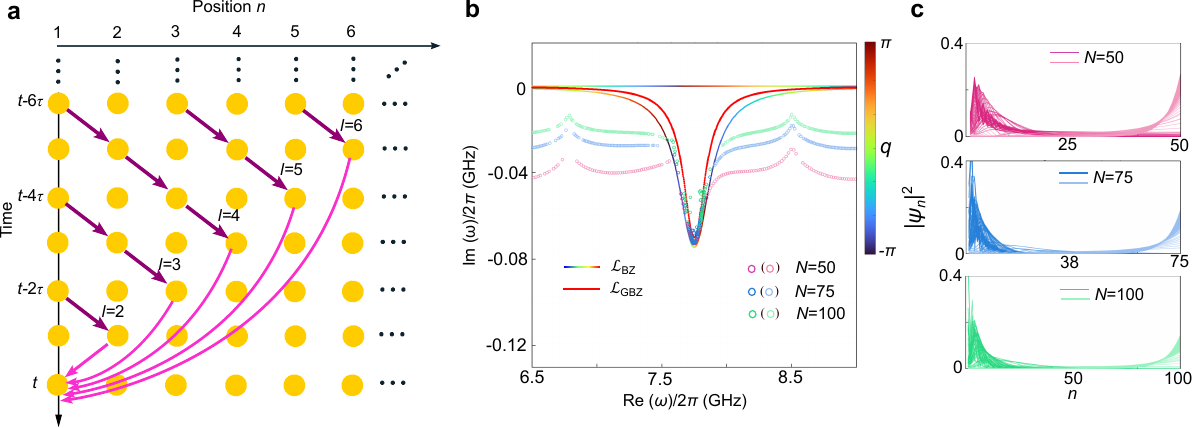}
    \caption{\textcolor{black}{Feedback-loop spatiotemporal picture and finite-size spectral evolution with system size \(N\). (a) Spatiotemporal representation of coherent returning loops under symmetric retardation (\(\tau_L=\tau_R=\tau, \Gamma_R \gg \Gamma_L\)). Yellow dots denote spatiotemporal lattice sites; purple arrows indicate the dressed right-cascaded propagation \(1 \to 2 \to \cdots \to l\), while pink arrows denote the direct leftward return \(l \to 1\). The coherent sum of these loops gives the feedback factor \(1/(1 - \beta z_L)\). (b) The discrete-spectrum evolution of the retarded system under ABC with system size \(N\) (\(\widetilde{\omega}_0/2\pi = 7.75 - i0.08\) GHz, \(\Gamma_L/2\pi =10^{-5}\), \(\Gamma_R/2\pi = 0.08\) GHz, \(\tau_L = \tau_R = \tau = 0.59\) ns): \(N=50\) (purple circle), \(75\) (blue circle), and \(100\) (green circle). The dark (light) color denotes the ABC eigenfrequencies \(\omega_{\mathrm{ABC}}\) inside (outside) the spectrum loop \(\mathcal{L}_{\mathrm{BZ}}\). With the increase of \(N\), the ABC spectrum gradually converges to the GBZ spectrum \(\mathcal{L}_{\mathrm{GBZ}}\) (red line) of GBZ, (c) The corresponding eigenstate spatial profiles of discrete spectrum under different system size \(N\).}}
\end{figure}

In the thermodynamic limit \((N\to\infty)\), \(|\beta_+|=|\beta_-|\), which is
precisely the condition defining the GBZ. Under the ABCs, there only exist multiple internal feedback. The eigenmode spatial profile is determined by
\(|\beta_-|=e^{(\omega_i+\Gamma_w)\tau_L}\).
When  \(\Gamma_0>\Gamma_w\), memory gain dominates the channel loss (\(\omega_i+\Gamma_w<0\)), we have \(|\beta_-|<1\), and thus
the retarded feedback fields are enhanced,
enabling the spatial profile to decay towards right side.

For a finite chain of length \(N\), boundary truncation must be treated
explicitly, because the second-order recurrence captures only the bulk
behavior. Substituting the ansatz \(a_m=\beta^m\) into the original
non-local equation gives
\(
R_n(\beta)
=\sum_{1\leq m<n}z_R^{n-m}\beta^m
= (z_R\beta^n-z_R^n\beta)/(\beta-z_R)
\)
and
\(
L_n(\beta)
=\sum_{n<m\leq N}z_L^{m-n}\beta^m
=z_L\beta^{n+1}/(1-z_L\beta)-z_L^{N-n+1}\beta^{N+1}/(1-z_L\beta)
\).
Then Eq.~(S58) becomes
\begin{equation}
\mathcal{E}_n(\beta)
=
\left[
A+i\Gamma_R\frac{z_R}{\beta-z_R}
+i\Gamma_L\frac{z_L\beta}{1-z_L\beta}
\right]\beta^n
-i\Gamma_R\frac{\beta}{\beta-z_R}z_R^n
-i\Gamma_L\frac{\beta^{N+1}}{1-z_L\beta}z_L^{N-n+1}
=0 .
\end{equation}
The first term is the bulk contribution. Thus, for \(\beta=\beta_{\pm}\), only
the boundary-truncation terms (i.e., the second and third terms) remain. For the finite-size eigenstate
\(a_n=c_1\beta_+^n+c_2\beta_-^n\),
the residual boundary terms impose
\begin{equation}
c_1\frac{\beta_+}{\beta_+-z_R}
+
c_2\frac{\beta_-}{\beta_--z_R}
=0,
\qquad
c_1\frac{\beta_+^{N+1}}{1-z_L\beta_+}
+
c_2\frac{\beta_-^{N+1}}{1-z_L\beta_-}
=0 .
\end{equation}
Eliminating \(c_1\) and \(c_2\) gives the finite-size quantization condition
\begin{equation}
\left| \frac{\beta_+}{\beta_-} \right|^N = \left| \frac{z_R - \beta_-}{z_R - \beta_+} \right| \left| \frac{1 - z_L \beta_+}{1 - z_L \beta_-} \right| = \left| \frac{(z_R - \beta_+)(z_R - \beta_-)}{(1 - z_L \beta_+)(1 - z_L \beta_-)} \right| \left| \frac{1 - z_L \beta_+}{z_R - \beta_+} \right|^2 .
\end{equation}
To further simplify this condition, define
\[
P(\beta)
=
(\beta-z_R)(1-z_L\beta)
\left[
A+i\Gamma_R\frac{z_R}{\beta-z_R}
+i\Gamma_L\frac{z_L\beta}{1-z_L\beta}
\right].
\]
This polynomial can be factorized as
\(
P(\beta)=C(\beta-\beta_+)(\beta-\beta_-)\) with
\(
C=z_L(-A+i\Gamma_L).
\)
Evaluating \(P(\beta)\) at \(\beta=z_R\) and \(\beta=1/z_L\), one obtains
\(
P(z_R)
=C(z_R-\beta_+)(z_R-\beta_-)
=i\Gamma_R z_R(1-z_L z_R),
\)
and
\(P\left(1/z_L\right)
=
C\left(1/z_L-\beta_+\right)
\left(1/z_L-\beta_-\right)
=i\Gamma_L\left(1/z_L-z_R\right)
\).
These relations lead to
\((z_R-\beta_+)(z_R-\beta_-)/[(1-z_L\beta_+)(1-z_L\beta_-)]=\Gamma_R z_R/(\Gamma_L z_L)
\).
Substituting this identity into Eq.~(S67), the ratio of the two roots satisfies
\begin{equation}
\left|
\frac{\beta_+}{\beta_-}
\right|
=
\left(
\frac{\Gamma_R}{\Gamma_L}
\left|
\frac{z_R}{z_L}
\right|
\left|
\frac{1-z_L\beta_+}{z_R-\beta_+}
\right|^2
\right)^{1/N}.
\end{equation}
When \(N\to\infty\), the condition \(|\beta_+|=|\beta_-|\) defines the GBZ,
implying that the discrete spectrum under the ABCs converges to the continuous
spectrum of the GBZ as the lattice size increases. As shown by the numerical results in
Fig.~S4(b), this convergence is clearly observed when the system size \(N\)
increases from \(50\) to \(100\). 

Furthermore, for a finite-size system under the ABCs, the eigenfrequencies near
\(\tilde{\omega}_0\) are predominantly governed by mode \(\beta_-\), as can be
deduced from Eq.~(S68). In the case of nearly unidirectional right coupling, where
\(
\beta_+\approx z_R(A-i\Gamma_R)/A
\),
we obtain
\begin{equation}
\left|
\frac{\beta_+}{\beta_-}
\right|
\approx
\left[
\frac{
\left|
2(1-z_L z_R)(\omega-\tilde{\omega}_0)
+i\left[\Gamma_L+\Gamma_R
+z_L z_R(\Gamma_R-\Gamma_L)
\right]
\right|^2
}
{4\Gamma_L\Gamma_R |z_L z_R|}
\right]^{1/N}.
\end{equation}
\textcolor{black}{As \(\omega\) approaches \(\widetilde{\omega}_0\), the modulus separation of two roots is reduced. Since \(|\beta_-| \approx |z_L|^{-1}\) is independent of the detuning \(\Delta = \omega_r - \omega_0\) (up to \(O(\Gamma_L/\Gamma_R)\)), whereas \(|\beta_+| \approx |z_R| |T_R|\) is minimized at \(\Delta = 0\) for \(\Gamma_0 > \Gamma_w\), the reduced modulus separation reflects a suppression of the right-cascaded propagation,} and the spatial profile
is primarily determined by the feedback-induced mode \(|\beta_-|\). Consequently,
in the regime of \(\Gamma_0>\Gamma_w\), the corresponding eigenstates
exhibit an anomalous left-localized spatial profile. Conversely, detuning from \(\widetilde{\omega}_0\) makes right-cascaded dynamics dominate, restoring the right-skin localization. This right-skin effect is weakened as the system size $N$ increases, as shown in Fig.~S4(c).

\begin{figure}[t]
    \centering
    \includegraphics[width=0.8\textwidth]{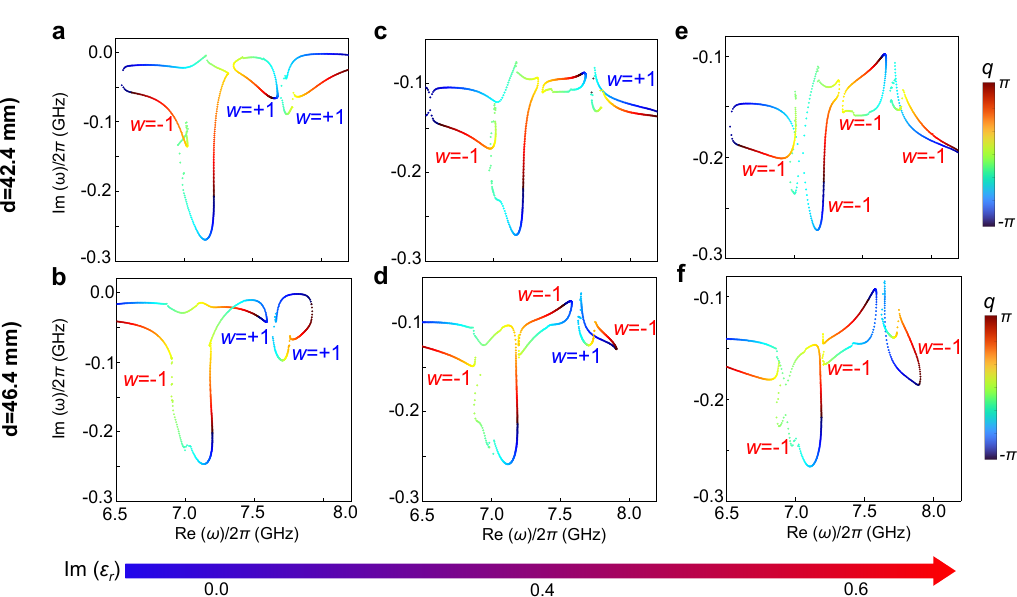}
     \caption{Simulated complex spectra \(\mathcal{L}_{\text{BZ}}\) of the microwave YIG-waveguide coupled system with different periods and waveguide intrinsic losses. The top (a, c, e) and bottom (b, d, f) rows correspond to the cases with periods of \(d=42.4 \text{ mm}\) and \(46.4 \text{ mm}\), respectively. From left to right columns, the waveguide intrinsic loss (\(\text{Im}(\varepsilon_r)\)) gradually increases while setting \(\text{Re}(\varepsilon_r)=12\): (a, b) \(\text{Im}(\varepsilon_r)=0\), (c, d) \(\text{Im}(\varepsilon_r)=0.4\), (e, f) \(\text{Im}(\varepsilon_r)=0.6\).}
\end{figure}
\section{Impact of air-channel coupling in microwave systems}
For the realistic microwave YIG cylinder-waveguide coupled system, the waveguide-mediated coupling coexists with the free-space coupling among the cylinders. At small inter-cylinder spacing, this reciprocal free-space coupling can deform the complex-spectrum loop generated by the mere waveguide-mediated coupling, giving rise to extra spectral loops.

Figure S5 shows the evolution of the complex spectra \(\mathcal{L}_{\text{BZ}}\) under different periods and waveguide losses. Because the free-space electromagnetic fields decay over distance, increasing the spatial period \(d\) substantially weakens this reciprocal air-channel coupling, thereby suppressing its interference with the primary waveguide-mediated nonreciprocal interaction. As observed when expanding the period from 42.4 mm to 46.4 mm, the spectral loops due to the free-space coupling are significantly suppressed, gradually shrinking and tending to vanish.

The waveguide loss can reduce the memory gain and even induce topological phase transition. In the realistic microwave system, we can control the imaginary part of permittivity of waveguide to gradually tune the waveguide loss. When \(\text{Im}(\varepsilon_r)\) increases from 0 to 0.6, the winding numbers of the two right-side loops gradually change from +1 to -1, while the two left-side loops preserve \(w=-1\), as depicted in Fig. S5. This is because that the channel dissipation is within the loss of the two dominated cylinder modes, as explained in the main text. Meanwhile, it should be noted that at \(\text{Im}(\varepsilon_r)=0.4\), the increased period can twist the rightmost loop to form a 8-shaped configuration with opposite winding numbers, which may be attributed to the reduced competition between the air and waveguide channels. Especially, the winding number of the right part of the spectrum at higher frequencies can be more easily reversed by the waveguide loss.

\textcolor{black}{
\section{Angle-selective excitation of the skin modes}
The eigenmodes of the NHSE can be excited by free-space \(s\)-polarized plane waves to facilitate light funneling. Crucially, this excitation process is angle-selective. For a finite-length system (\(N = 9\)) with a period of \(d = 46.4\) mm under OBC, the simulated field distributions within the waveguide as a function of the incident angle \(\theta\) at the frequency \(7.76\) GHz are illustrated in Figs. S6(a) and (c). In the waveguide-lossless scenario [Fig. S6(a)], the left skin mode is prominently excited near \(\theta = -20^\circ\), whereas the right skin effect emerges near \(\theta = 12^\circ\) in the channel-loss scenario [Fig. S6(c)]. This angular selectivity stems from the momentum-matching condition. The corresponding Fourier spectra of the eigenmodes near \(7.76\) GHz are calculated and plotted in Figs. S6(b) and S6(d) for the lossless and lossy cases, respectively. Notably, the spectral peaks are located at \(-19.2^\circ\) and \(9.4^\circ\) for the lossless-waveguide case, and at \(-18.8^\circ\) and \(9.1^\circ\) for the lossy-waveguide case. Although the optimal excitation angles closely align with these Fourier peaks, the coexistence of multiple eigenmodes near the excitation frequency dictates that the resulting spatial profile is a linear combination of these adjacent modes. Consequently, the interference among these superposed eigenmodes can obscure the overall localization feature. This explains why the skin localization is not obvious around \(9.4^\circ\) in Fig. S6(a) and \(-18.8^\circ\) in Fig. S6(c).}

\begin{figure}[t]
    \centering
    \includegraphics[width=0.6\linewidth]{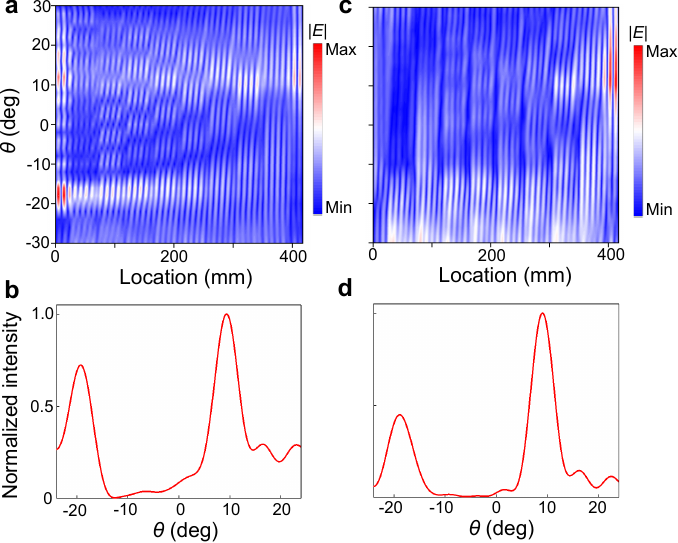}
    \caption{\textcolor{black}{Angle-selective excitation of the skin modes in the finite-length non-Markovian system (\(N = 9\)) with a period of \(d = 46.4\)mm. (a, c) Simulated field distributions within the waveguide as a function of the incident angle \(\theta\) at the frequency \(7.76\) GHz for (a) the lossless waveguide (\(\varepsilon_r = 12\)) and (c) the lossy waveguide (\(\varepsilon_r = 12 + 0.4i\)). (b, d) Corresponding Fourier spectra based on the respective eigenmodes near \(7.76\) GHz for the system (b) without and (d) with waveguide losses.}}
\end{figure}

\section{Experimental measurements}
\textbf{External magnetic-field bias:} To generate the external magnetic field environment required for YIGs, we place a pair of N35 magnets, each with a diameter of 40 mm and a thickness of 5 mm, on both sides of the YIG cylinder, as shown in Fig. S7(a). The simulation results show that when the distance between the two magnets is 30 mm, the static magnetic field between them can uniformly keep 0.1 T. For the array system, we combine 10 pairs of N35 magnets to guarantee the wide-range uniform magnetic bias near 0.1 T.

\textbf{Dielectric waveguide to metallic waveguide transition:} Following our previously reported technique \cite{cheng2023directional,xue2025achieving}, we employ a dielectric-waveguide-to-metallic-waveguide (DWG-to-MWG) transition to connect the components. As shown in Fig. S7(b), the central DWG (\(\varepsilon_r = 12\), \(\mu_r = 1\)) section has a length of \(L = 424\) mm, a width of \(a = 4\) mm, and a thickness of \(b = 13\) mm. At both ends, the DWG transitions into two standard WR137 metallic waveguide launchers (MWG 1 and MWG 2) via tapered structures with a length of \(l_t = 15\) mm. For experimental verification, the metallic waveguide launchers are connected to Ports 1 and 2 of a vector network analyzer (VNA). The measured reflection coefficients (\(S_{11}\) and \(S_{22}\)) and transmission coefficient (\(S_{21}\)) are plotted in Fig. S7(b) over the frequency range of 7.0 to 8.2 GHz. The experimental results show that the transmission coefficient (\(S_{21}\)) is around -2.5 dB across the desired frequency band, demonstrating that the energy is efficiently coupled from Port 1 to Port 2 with favorable transmission performance. Concurrently, the reflection coefficients (\(S_{11}\) and \(S_{22}\)) are well-suppressed below -10 dB, meaning that less than 10\% of the incident power is reflected back. These results confirm that the tapers effectively ensure a smooth transition and stable impedance matching between the dielectric waveguide and the metallic waveguide launchers.

\textbf{Waveguide-loss control:} To perform the experiment investigating the skin effect in lossy dielectric waveguides, we introduce specific losses by attaching microwave absorbers to both sides of the dielectric waveguide as shown in Fig. S7(c). The absorber used is a thin, flexible, high-loss, electrically non-conductive silicone absorber (Laird Eccosorb BSR / MFS), which is designed for frequencies ranging from 6 GHz to 35 GHz, exhibiting high magnetic loss and high temperature stability. According to the manufacturer, at our frequency of interest (7.76 GHz), the attenuation of the absorber material reaches approximately 35 dB/cm. To quantify the introduced loss, we first performed numerical simulations. Specifically, the complex permittivity of the waveguide section was modified from 12 to \(12 + 0.4i\), resulting in simulated \(S\)-parameters of \(|S_{11}|=-12.12\) dB and \(|S_{21}|=-20.31\) dB (noting that before introducing the loss, \(|S_{11}|<-10\) dB and \(|S_{21}|\simeq-3\) dB). For experimental verification of the simulated result, we attached different thicknesses of the absorbing material onto both sides of the dielectric waveguide. The results show that when the thicknesses are set to \(t_1=1.28\) mm and \(t_2=1.92\) mm with a length of \(l_{abs}=305\) mm and a width of \(w_{abs}=4\) mm, the simulation results closely match the measured results of \(|S_{11}|=-12.31\) dB and \(|S_{21}|=-24.9\) dB. This confirms the successful and controlled introduction of loss in the waveguide system.
\begin{figure}
    \centering
    \includegraphics[width=0.8\textwidth]{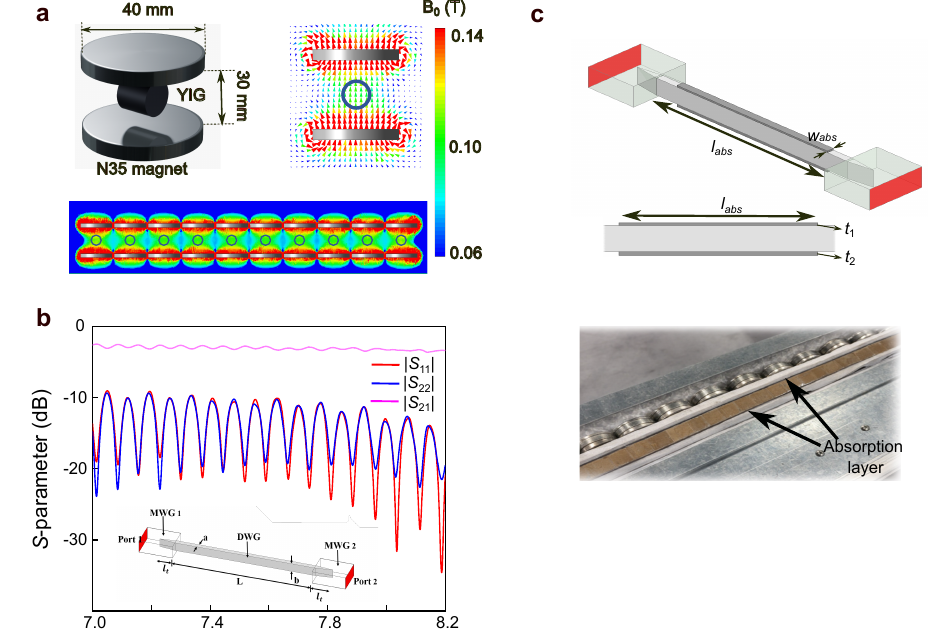}
    \caption{ (a) A pair of magnets providing uniform external magnetic bias. (b) Measured S-parameters of the proposed dielectric waveguide (DWG) to metallic waveguide (MWG) transition. The inset shows the schematic configuration and geometric parameters of the transition structure. (c) Schematic of the externally attached absorbing patches used to control the waveguide channel loss. The bottom panel shows a photograph of the actual experimental setup.}
\end{figure}
\end{document}